\documentclass[
 aps, pra,
 amsmath,amssymb,
 11pt,
 final,
tightenlines,
 twoside,
 twocolumn,
 nofloats,
nofootinbib,
 superscriptaddress,
showkeys,
showkeywords,
 ]
{revtex4-2}

\usepackage[T2A]{fontenc}
\usepackage[utf8x]{inputenc}
\usepackage[english]{babel}
\usepackage{graphicx}
\usepackage{dcolumn}
\usepackage{bm}
\usepackage{xcolor}
\usepackage{longtable}
\usepackage{afterpage}

\newcommand{\hii}{H\,{II}}

\newcommand{\kms}{km\,s$^{-1}$}

\newcommand{\micron}{$\mu$m}
\input{maik.rty}

\setcitestyle{authoryear,round}
\def\squareforqed{\hbox{\rlap{$\sqcap$}$\sqcup$}}

\def\sq{\ifmmode\squareforqed\else{\unskip\nobreak\hfil
\penalty50\hskip1em\null\nobreak\hfil\squareforqed
\parfillskip=0pt\finalhyphendemerits=0\endgraf}\fi}

\def\arcsec{\hbox{$^{\prime\prime}$}}

\def\utw{\smash{\rlap{\lower5pt\hbox{$\sim$}}}}

\def\udtw{\smash{\rlap{\lower6pt\hbox{$\approx$}}}}

\def\diameter{{\ifmmode\mathchoice
{\ooalign{\hfil\hbox{$\displaystyle/$}\hfil\crcr
{\hbox{$\displaystyle\mathchar"20D$}}}}
{\ooalign{\hfil\hbox{$\textstyle/$}\hfil\crcr
{\hbox{$\textstyle\mathchar"20D$}}}}
{\ooalign{\hfil\hbox{$\scriptstyle/$}\hfil\crcr
{\hbox{$\scriptstyle\mathchar"20D$}}}}
{\ooalign{\hfil\hbox{$\scriptscriptstyle/$}\hfil\crcr
{\hbox{$\scriptscriptstyle\mathchar"20D$}}}}
\else{\ooalign{\hfil/\hfil\crcr\mathhexbox20D}}%
\fi}}

\begin{document}

\selectlanguage{english}


\title{Chemical differentiation and gas kinematics around massive young stellar objects in~RCW\,120}

\author{\firstname{K.~V.}~\surname{Plakitina}\footnote{plakitina.kv@inasan.ru} }
\affiliation{Institute of Astronomy, Russian Academy of Sciences, Pyatnitskaya str. 48, Moscow 119017, Russia}

\author{\firstname{M.~S.}~\surname{Kirsanova} }
\affiliation{Institute of Astronomy, Russian Academy of Sciences, Pyatnitskaya str. 48, Moscow 119017, Russia}
\affiliation{Astro Space Center, Lebedev Physical Institute, Russian Academy of Sciences, 117997, 84/32 Profsoyuznaya Str., Moscow, Russia}


\author{\firstname{S.~V.}~\surname{Kalenskii} }
\affiliation{Astro Space Center, Lebedev Physical Institute, Russian Academy of Sciences, 117997, 84/32 Profsoyuznaya Str., Moscow, Russia}

\author{\firstname{S.~V.}~\surname{Salii} }
\affiliation{Institute of Natural Sciences and Mathematics, Ural Federal University, 19 Mira Str., 620075 Ekaterinburg, Russia}

\author{\firstname{D.~S.}~\surname{Wiebe} }
\affiliation{Institute of Astronomy, Russian Academy of Sciences, Pyatnitskaya str. 48, Moscow 119017, Russia}


\begin{abstract}

We present results of a spectral survey towards a dense molecular condensation and young stellar objects (YSOs) projected on the border of the \hii{} region RCW\,120 and discuss emission of 20 molecules which produce the brightest lines. The survey was performed with the APEX telescope in the frequency range 200 -- 260~GHz. We provide evidences for two outflows in the dense gas. The first one is powered by the RCW\,120~S2 YSO and oriented along the line of sight. The second outflow around RCW\,120~S1 is aligned almost perpendicular to the line of sight. We show that area with bright emission of CH$_3$OH, CH$_3$CCH and CH$_3$CN are organised into an onion-like structure where CH$_3$CN traces warmer regions around the YSOs than the other molecules. Methanol seems to be released to the gas phase by shock waves in the vicinity of the outflows while thermal evaporation still does not work towards the YSOs. We find only a single manifestation of the UV radiation to the molecules, namely, enhanced abundances of small hydrocarbons CCH and c-C$_3$H$_2$ in the photo-dissociation region. 

\end{abstract}

\maketitle

\section{Introduction}\label{sec:intro}

Formation of massive stars and their feedback on the surrounding interstellar medium (ISM) have attracted the attention of researchers since the second half of the XXth century \citep[see e.~g. reviews by][]{2007ARA&A..45..481Z, 2018ASSL..424..119F}. These stars are rare but they bring powerful ultraviolet (UV) radiation and momentum to the ISM, changing its physical and chemical state. It has been suggested that the feedback can lead to formation of new generations of stars via a process of triggered/stimulated collapse in dense envelopes of \hii{} regions \citep{1977ApJ...214..725E}. This can be established with more certainty in studies of large-scale structures (of the order of kiloparsecs) in external galaxies because it is possible to distinguish separate generations of stars and ionized envelopes out there \citep[e.~g.][]{Egorov_2014,Egorov_2017}. However, on the scale of Galactic star-forming regions, it can be complicated to prove that the formation of a certain stellar population was triggered \citep[][]{Preibisch_1999, 2022A&A...667A.163M, 2023MNRAS.522.1288B, 2023arXiv230207853R}. In many cases possible triggering is claimed on the basis of mutual location of embedded young stellar clusters, dense molecular gas and ionized gas around young massive stars but it is certainly not sufficient \citep[see][]{2015MNRAS.450.1199D}.

At the times of large-scale surveys and widespread correlation analysis, it is tempting to distinguish between different star formation processes on large maps using e.~g. known physical conditions in star-forming regions and line intensities or molecular column densities/abundances. For example, \citet{2017AA...599A..98P} demonstrated that N$_2$H$^+$ and CH$_3$OH emission lines are the most reliable tracers of  the high density molecular gas through the spectral lines at 3~mm, see e.~g.~\citet{2017AA...599A..98P}, unlike other widely used tracers such as lines of HCO$^+$ or HCN. \citet{2018A&A...610A..12B} were able to separate between UV-illuminated and UV-shielded gas using several isotopes of CO and HCO$^+$ in the same molecular cloud. Based on the analysis of molecular column densities, \citet{2021MNRAS.507.3810K} have shown that molecular clouds around \hii{} regions S\,235 and S\,235\,A reveal a molecular content typical for dark star-forming clouds without signatures of UV~irradiation. Large-scale images in the near-infrared allows distinguishing between envelopes created by the main sequence massive stars and by young stellar objects embedded into molecular clouds \citep[e.~g.][and many others]{2009MNRAS.398.1368R, 2021ApJ...919...27K, 2023MNRAS.518.2320D, 2023AstBu..78..372K}.

 The RCW\,120 PDR has a ring-like shape; it has been formed in a flattened molecular cloud \citep[][]{Anderson_2015, Kirsanova_2019, Zavagno_2020}. 
Previously, \citet{Figueira_2017} extracted a number of compact sources at the edge of \hii{} region RCW\,120, using Herschel observational data of the RCW\,120 region. Among these sources, they detected objects characterised by massive envelopes, which are associated with massive YSOs. Subsequent studies by \citet{Figueira_2020} cast doubt on the triggering scenario for RCW\,120 and suggested that the YSOs around the \hii{} pre-dated the expansion of the \hii{} region. \citet{2023MNRAS.520..751K} have detected the dense layer shocked by the expanding \hii{} region, found the YSOs outside of this layer, thus, confirming the latter result.

In spite of the studies mentioned above, the question about the influence of the expanding \hii{} region on the present star-formation process in RCW\,120 is still open. This question is interesting because of its possible influence on the mass function or time scale for the star formation \citep[e.~g. see discussion by][]{Luisi_2021}. Recently \citet{Kirsanova_2021} analysed emission of CH$_3$CN and CH$_3$OH molecules in two YSOs around RCW\,120. They estimated abundances of these molecules and gas physical parameters in both YSOs and found that a massive YSO RCW\,120~S2 \citep{Figueira_2017} demonstrates the beginning of a warm-up phase preceding the establishing of the hot gas chemistry. A less massive YSO RCW\,120~S1 might be at an even less evolved evolutionary stage.  We study gas temperature, molecular abundances and gas kinematics around the YSOs to find out if the vicinity of the \hii{} region influences these sources now.

\section{Observational data}\label{sec:observations}

\begin{figure}[h!]
\includegraphics[width=\columnwidth]{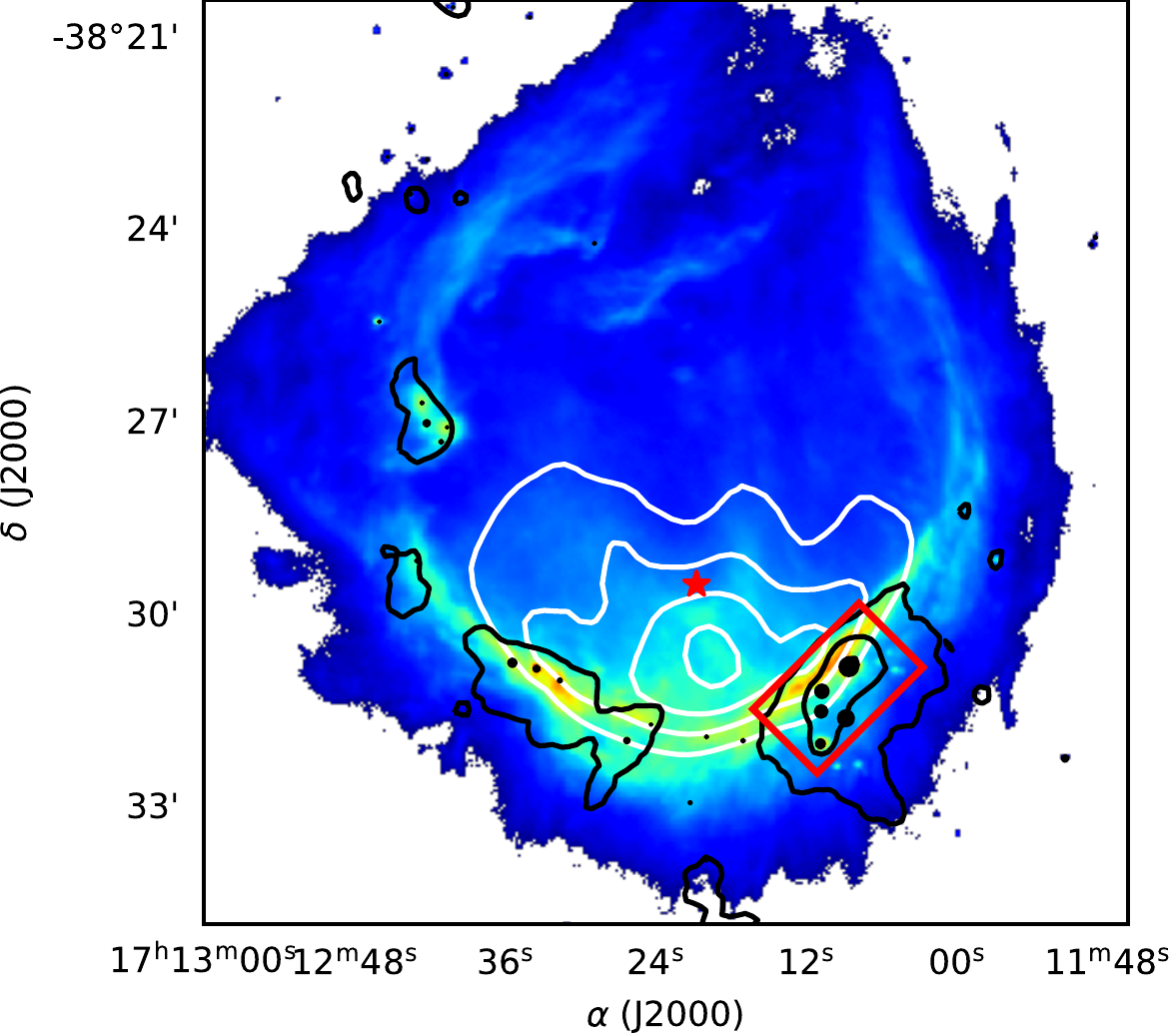}
\caption{Image of RCW\,120 at 70\micron{}. Colour scale shows emission of neutral material around the ionized gas. The position of the ionizing star of RCW~120 is marked as a red star. The 843~MHz radio continuum emission \citep{1999AJ....117.1578B} corresponding to free-free emission of the ionized gas is shown with white contours linearly spaced from from 0.1 to 0.4 Jy/beam. The 870\,$\rm \mu$m \citep{2009A&A...496..177D} contours for 0.4, 2.0 and 10.0 Jy/beam are shown in black. Black circles show the locations of compact sources \citep{Figueira_2017}. The sizes of the circles depend linearly on the source masses here and on the following figures. The area, observed in this study, is shown by the red rectangle.}
\label{fig:observed}
\end{figure}

In this study, we consider a dense molecular condensation located right behind the south-west border of the RCW~120 \hii{} region along with several YSOs, S1, S2, S9, S10, and S39, embedded into this condensation. The observed area is shown in Fig.~\ref{fig:observed}.

\subsection{APEX observations}

The observations were carried out using the Atacama Pathfinder EXperiment telescope (APEX) in Chile \citep{2006A&A...454L..13G} on 16--23 of September 2021, as project O-0108.F-9313-2021 (PI: Kirsanova M.~S.) within the Swedish operated share. The heterodyne receiver used was the nFLASH230. This sideband separating receiver simultaneously covers 2 bands of 8~GHz in both polarizations.
Using nFlash230 tunned to 217~GHz in the USB setup, the observations covered bands from 199~GHz to 207~GHz and from 215~GHz to 223~GHz.  The LSB setup was tuned to 241~GHz, therefore our observations spanned bands from 238~GHz to 246~GHz and from 258~GHz to 262~GHz.
The spectral resolution in the FFT spectrometer used was about 0.3~\kms. The spatial resolution at 230~GHz was 26\arcsec, corresponding to 0.17~pc at the distance of RCW\,120 of 1.34~kpc~\citep[see distance to the ionizing star in][]{Russeil_2003}. We did two OTF maps of 140\arcsec{}~by~85\arcsec{} rotated by 45$^\circ$\ in the equatorial system (with the long side extending from southeast to northwest) with a dump-time of 1~sec. The rows/columns in the map were alternatively observed along the x- and y-directions with a data dump time of 1.0~sec and a step of 1/3 of the beam size.

The data were calibrated to antenna temperature in real-time using the standard {\it apexOnlineCalibrator} package, but we later additionally applied factors of $\eta_{\rm mb} = 0.79$\footnote{https://www.apex-telescope.org/telescope/efficiency/ index.php}, to arrive at the main beam temperature scale.

The gridding of the data and the baseline correction were performed using the CLASS package from the GILDAS\footnote{\url{https://www.iram.fr/IRAMFR/GILDAS}} software. Further analysis was done with the Astropy package~\citep{astropy:2013, astropy:2018}, and APLpy~\citep[][]{aplpy2012, aplpy2019} was used for representation. Final fits-cubes were converted to a common grid with the step size of 10\arcsec{} in x- and y-directions. Typical noise level ($1\sigma$) of the re-gridded data is 20--40~mK. The entire program required $\approx 10$~hours of the observing time (including telescope and calibration overheads).

\subsection{Archival data}

In addition to our own observations of molecular lines, we used prepared data-cubes with integrated intensity of SiO(2--1) emission line from MALT90 Survey \citep[][]{2013PASA...30...57J}.

In order to estimate column densities of hydrogen nuclei $N({\rm HI+H_2})$ and dust temperature $T_{\rm dust}$, we used Hi-GAL images \citep{2016AA...591A.149M} at 70, 160, 250, 350, and 500 $\mu$m, downloaded from the SSDC web-site\footnote{https://tools.ssdc.asi.it/HiGAL.jsp}, along with the ATLASGAL data \citep{2009A&A...504..415S} at 870~$\mu$m\footnote{https://atlasgal.mpifr-bonn.mpg.de/cgi-bin/ATLASGAL\_DATABASE.cgi}.

\section{Methods} 

We study the molecular condensation bordering the RCW\,120~PDR. As it was shown before \citep[e.~g.][]{Kirsanova_2019, Figueira_2020, Kirsanova_2021, Kabanovic_2022, 2023MNRAS.520..751K}, an LTE approximation gives reasonable results for this molecular region. Therefore, we use this approximation to analyse emission of all detected molecules except for methanol (see below). For those molecules, where we have three or more transitions, we built rotational diagrams to determine rotational temperatures. For all other cases, we assumed that the excitation temperature is equal to the dust temperature, determined from the analysis of far-IR {\it Herschel} and ATLASGAL data.

\subsection{Rotational diagram technique}\label{sec:rotation_diagrams}

Such molecules as methyl acetylene (CH$_3$CCH) and methyl cyanide (CH$_3$CN) are frequently used as ``thermometers'' for warm and dense gas \citep[e.g.][]{Kalenskii_2000,Giannetti_2017,Andron_2018,Calcutt_2019,Brouillet_2022}. Both species are symmetric top molecules. The rotational transitions of  CH$_3$CCH and CH$_3$CN are described by two quantum numbers: the total angular momentum ($J$) and its projection along the symmetry axis ($K$). The molecular transitions of these molecules cover a broad range of upper energy levels. The lines $J\rightarrow J-1$ with the same $K$ numbers are closely spaced in frequency and can be observed sumultaneously. 

We apply a technique of rotational diagrams in order to determine excitation temperatures and column densities of CH$_{3}$CCH, CH$_3$CN, CH$_3$OH, and SiO molecules. \cite{Askne1984} have shown that CH$_{3}$CCH is a relevant species to probe kinetic temperature in molecular gas as it has a relatively low dipole moment, namely 0.784~D \citep{Burrell_1980}. CH$_3$CN is thermalized only in regions of high density, as it has a dipole moment of 3.92~D \citep{Gadhi_1995}. In addition to building rotational diagrams to determine methanol column densities over the entire map, we also applied a non-LTE method to estimate physical parameters using methanol line emission in the directions of the YSOs as non-LTE effects can be important \citep[][]{2016Kalenskii}. Bright methanol emission towards the YSOs allows us using the non-LTE method and confining physical parameters there.

For the analysis, we followed \citet{Goldsmith1999}, who summarised the use of the rotational diagram technique. This method refers to a plot of the natural logarithm of the column density of molecules in the upper state $N_{\rm u}$:
\begin{equation}
   \ln \left(\frac{N_{\rm u}^{\text {thin }}}{g_{\rm u}}\right)=\ln \left(\frac{N_{\text {tot }}^{\text {thin }}}{Q\left(T_{\text {rot }}\right)}\right)-\frac{E_{\rm u}}{k_{\rm B} T_{\text {rot }}},
   \label{eq_N_tot}
\end{equation}
where $T_{\text{rot}}$ is the rotational temperature, $Q$ is the partition function for the specific value of $T_{\text{rot}}$, $g_{\rm u}$ is upper state degeneracy, $E_{\rm u}$ is the energy of the upper level, $k_{\text{B}}$ is the Boltzmann constant. The column density of molecules $N_{\rm u}^{\rm thin}$ can be found as:
\begin{equation}
	N_{\rm u}^{\text {thin }}=\frac{8 \pi k_\text{B} \nu^2 W}{hc^3 A_{\rm ul}}, 
	\label{eq_N_u}
\end{equation}
where $\nu$ is the rest frequency, $W=\int T_{\text{mb}} d V$ is the integrated line intensity, $h$ is the Planck constant, $c$ is the speed of light, $A_{\rm ul}$ is the Einstein coefficient of spontaneous emission from upper to lower level. To plot the rotational diagram, we considered only lines with the signal-to-noise ratio $>3$. 

If the lines are optically thin and corresponding transitions have approximately the same excitation temperature, a plot of $\ln(N_{\rm u}^{\text {thin}}/g_{\rm u})$ versus $E_{\rm u}/{\rm k}_{\text{B}}$ appears as a straight line and we are able to derive rotational temperature, which is inversely proportional to the tangent of its slope to the x-axis. The total column density of the molecule $N_\text {tot}$ can be found then from the point where the plotted line intercepts y-axis as $\ln(N_\text {tot}^{\text{thin }}/Q(T_{\text{rot}})$.
Partition function $Q(T_{\text{rot}})$ depending on $T_{\text{rot}}$ was found by interpolating data provided in  The Cologne Database for Molecular Spectroscopy, CDMS, \citet{2001AA...370L..49M} \footnote{\url{https://cdms.astro.uni-koeln.de}} to the derived rotational temperature. We used the $Q$ values including both $A$ and $E$ types of CH$_3$CCH, CH$_3$CN and CH$_3$OH molecules.

All the uncertainties were estimated using the error propagation formalism. 

\subsection{Molecular column densities in LTE}\label{sec:LTE_CD}

For those molecules, where we could not estimate the excitation temperature, we followed \cite{Mangum2015} to determine molecular column densities.

First, we calculated column densities using optically thin approximation:
\begin{equation}
\begin{aligned}
	N_{\text{tot}}^{\text{thin}}= & \left(\frac{3 {\rm h}}{8 \pi^3 S \mu^2 R_{\rm i}}\right)\left(\frac{Q\left(T_{\text {rot }}\right)}{g_{\rm u}}\right) \frac{\exp \left(\frac{E_{\rm u}}{k_\text{B} T_{\text{ex}}}\right)}{\exp \left(\frac{h \nu}{k_\text{B} T_{\text{ex}}}\right)-1} \\
	& \times \frac{1}{\left(J_\nu\left(T_{\text{ex}}\right)-J_\nu\left(T_{\text{bg}}\right)\right)} \int \frac{T_{\text{mb}} d V}{f},
\end{aligned}
\end{equation}
where line strength $S=\frac{J_u}{2 J_u+1}$ for linear molecules and $S=\frac{J_u^2-K^2}{J_u(2 J+1)}$ for symmetric top molecules, $\mu$ is the dipole moment taken from the CDMS, $R_i$ are the relative hyperfine intensities.  The partition function $Q$ was calculated for linear molecules ($Q \simeq \frac{k T_{\text{ex}}}{\rm h B}+\frac{1}{3}$) and interpolated using data presented in the CDMS for symmetric and asymmetric rotor molecules, $g_{\rm u}$ is the total degeneracy for an energy level of a transition, $T_{\text{ex}}$ is the excitation temperature, $J_\nu(T) \equiv \frac{h \nu}{k_\text{B}} / \left( \exp \left(\frac{h \nu}{k_\text{B} T}\right)-1 \right)$ is the equivalent intensity of a black body at temperature $T$. We assume the filling factor $f = 1$ and $T_\text{bg} = 2.7$~K hereinafter.

By conducting observations, spanning a wide frequency range, we were able to detect both optically thick lines and lines from the isotopologues. Given low abundances of isotopologues, it is reasonable to assume that the associated lines can be considered as optically thin, which allows estimating optical depth of the main isotope $\tau_{\nu}$ from the peak main-beam brightness temperatures $T_{\text{mb, thin}}$ and $T_{\text{mb, thick}}$ of optically thin and optically thick lines, respectively:
\begin{equation}
\frac{T_{\text{mb, thin}}}{T_{\text{mb, thick}}}=\frac{1-\exp \left(-\tau_{\nu} / r\right)}{1-\exp \left(-\tau_{\nu}\right)},
\label{eq: tau}
\end{equation}
where $r$ is isotope ratio, which depends on the galactocentric distance of the object \citep[we used values summarized by][]{Wilson1999}. The equation was solved by applying the {\it fsolve} function from the Python module scipy.optimize\footnote{https://docs.scipy.org/doc/scipy/reference/generated/ scipy.optimize.fsolve.html}. Using the same Python function and assuming that the medium is uniform, we obtained excitation temperature $T_{\text{ex}}$ as follows:
\begin{equation}
T_{\text{mb, thick}}=\left(J_\nu\left(T_{\text{ex}}\right)-J_\nu\left(T_{\text{bg}}\right)\right)\left(1-\exp \left(-\tau_\nu\right)\right).
\end{equation}
Based on the comparable collision cross sections we made the assumption that all isotopologues share the same excitation temperature. Using this assumption, we determined column densities of CO (using $^{13}$CO and C$^{18}$O isotopologues) and CS (combining with C$^{34}$S) molecules. Isotopic ratio $r = 50.1$ was applied for $^{13}$C/$^{12}$C and $r = 440.5$ for $^{16}$O/$^{18}$O, according to the galactocentric distance of RCW\,120 \citep[see ][]{Wilson1999}. For the $^{32}$S/$^{34}$S ratio we used $r = 22$. This ratio, as demonstrated by  \cite{Frerking_1980}, remains terrestrial and does not vary with distance from the galactic centre. For all other molecules we used dust temperature $T_{\rm dust}$ as $T_{\text{ex}}$.

Finally, considering the optical depth of the lines of the main isotopologues we can introduce correction for the optical depth to the column density ($N_{\text{tot }}^{\text{thick}}$):
\begin{equation}
N_{\text{tot }}^{\text{thick}}=N_{\text{tot}}^{\text{thin}} \frac{\tau}{1-\exp (-\tau)}.
\end{equation}

For the analysis of deuterated molecules and species lacking available isotopologue observations, we adopted an alternative approach. Considering their emission as optically thin, we used equations (\ref{eq_N_u}) and (\ref{eq_N_tot}) to calculate their column densities.

In order to calculate column densities of both ortho+para H$_2$CO and c-C$_3$H$_2$ molecules, we used a ratio of nuclear spin-weights 3:1 for ortho:para. To derive total column density of c-C$_{3}$H$_{2}$ we took into account that the ortho state is 1.63 cm$^{-1}$ above the ground.

\subsection{Non-LTE analysis of methanol}

Physical conditions towards the YSOs were estimated by 22 detected methanol emission lines in the LVG approximation. We used pre-calculated database of the population numbers for quantum energy levels of methanol presented by \cite{Salii_2018}. Here we describe the main features of the database. The population numbers for the methanol energy levels were calculated for the model of methanol excitation in typical conditions of star-forming regions. Namely, gas kinetic temperature ($T_{\rm k}$), molecular hydrogen number density ($n_{\rm H_2}$), specific methanol column density ($N/\Delta V$), and methanol abundance ($N/N_{\rm H_2}$) were varied within the ranges of the database (from 10 to 600 K for T$_{\text{k}}$ , from 3.0 to 9.0 for $\lg(n_{\rm H_2})$, from 7.5 to 14.0 for $\lg(N_{\rm CH_3OH}/\Delta V$), from -9.0 to -5.5 for $\lg(N_{\rm CH_3OH}/N_{\rm H_2})$ and from 1 to 5~\kms{} for FWHM). The model separately describes energy levels of A and E-type of methanol in the ground and torsionally excited levels with v$_{\rm t}$ up to~2. Quantum numbers of the considered states reach $J\leq22$ and $|K|\leq9$ with the upper state energies $E_{\rm up}=1015.5$ and 1020.2~cm$^{-1}$ for the A and E-type, respectively, see \citet{2005MNRAS.360..533C}. In order to consider dust emission within the region, we applied additional coefficients to the methanol optical depth and source function as described by \citet{2004ApJ...609..231S}. Collisions with hydrogen and helium were calculated using the coefficients from \citet{2010MNRAS.406...95R} and \citet{2010MNRAS.403.2033R}, respectively. The database contains the population numbers calculated for three values of the methanol line widths: 1, 3 and 5\,\kms. Therefore, to use the database, we fixed widths of all the methanol lines at the position of each YSO under study, namely ${\rm FWHM} = 5$~\kms{} for RCW\,120~S1 and S2, and 3~\kms{} for all the other YSOs (S9, S10, and S39). As it can be seen below from the spectra, our choice is reasonable for the considered YSOs. The beam filling factor $f$ was varied from 10 to 100\% with a step of 10\%. For all sources, the values of $T_{\text{mb}}$ obtained from the single Gaussian fitting were used. For all sources, we ensured that the intensities of unregistered lines do not exceed 1~$\sigma$.

\subsection{Hydrogen column density and dust temperature}

All the data about far-IR dust continuum emission data have been convolved to the resolution of the 500~$\mu$m {\rm Herschel} SPIRE map and re-gridded to the ATLASGAL image. Then, at each pixel we fitted observed intensities to the modified black body to get dust surface densities and temperatures, assuming that the observed intensity is given by
\begin{equation}\label{eq:tdust}
I_\lambda = B_\lambda(T)(1-e^{-\tau_\lambda})\approx B_\lambda(T)\tau_\lambda = B_\lambda(T)\Sigma\kappa_\lambda,
\end{equation}
where $B_\lambda(T)$ is the Planck function, $\tau_\lambda$ is the optical depth, $\Sigma$ is the dust surface density, $T$ is the dust temperature
and $\kappa_\lambda$ is the dust opacity. We made a usual assumption that
\begin{equation}
\kappa_\lambda=\kappa_0\left(\frac{\lambda_0}{\lambda}\right)^\beta.
\end{equation}
where $\kappa_0=7$~cm$^2$g$^{-1}$, $\lambda_0=450\,\mu$m, $\beta=2.2$, in line with estimates, presented by \citet{2011ApJ...728..143S}. The obtained surface density was converted into H nuclei column density, assuming the distance to RCW\,120 of 1.34~kpc. Finally, $N({\rm HI+H_2}) = 100\Sigma/(1.4m_{\rm H})$, where 100 is the assumed gas-to-dust mass ratio~\citep{1955ApJ...121..559L, 1978ApJ...224..132B}. While the ratio can be different in \hii{} regions due to radiation pressure \citep[see e.~g.][]{2011ApJ...732..100D, 2015MNRAS.449..440A, 2017MNRAS.469..630A, 2023MNRAS.526.5187K}, we study molecular clouds, therefore, use this value. 

\subsection{Kinematics of molecular gas}\label{sec:methodkinem}

In order to study gas kinematics, we have searched for high-velocity wings of the SiO~(5--4) and CH$_3$OH~(5$_{0, 5}$--4$_{0, 4}$) emission lines. These molecules are usually used to detect outflows in the vicinity of hot molecular cores \citep[e.~g.]{Sanchez-Monge2013,Codella_2013,Araya2008}. The molecular condensation under the study has a north-west -- south-east velocity gradient $\approx -1$\kms{} \citep[see][]{Kirsanova_2019, 2023MNRAS.520..751K}. Therefore, we have to take into account the gradient studying outflows by wings of spectral lines around the YSOs. To isolate the wings on the emission maps, we performed Gaussian fitting of the lines in each pixel. The emission falling outside the threshold, equal to the peak velocity~$\pm$~FWHM/2, were identified as blue and red intervals. The integrated emission falling inside the FWHM interval is considered as a bulk of the emission in each pixel and is called the middle interval below.

\section{Results} 

\subsection{Detected molecules}

\begin{figure*}
    \includegraphics[width=2\columnwidth,]{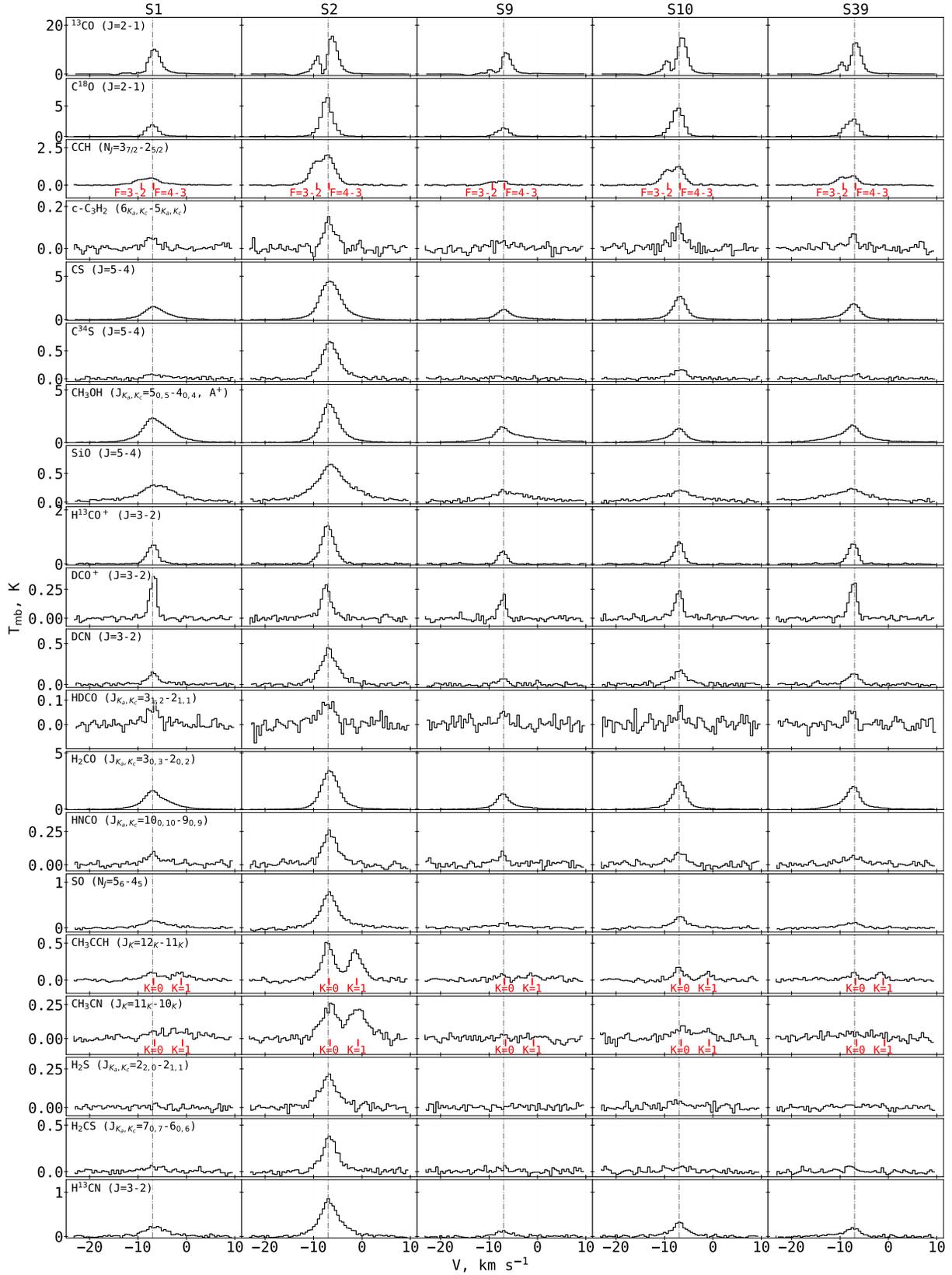}
    	\caption{Selected spectral lines of the twenty detected molecules towards five YSOs: S1, S2, S9, S10, S39. The dashed grey line marks $V_\text{LSR} = -7$~km\,s$^{-1}$ } . 
    	\label{fig:detectedlines}
\end{figure*}

The molecular condensation under the study has a common border with the RCW\,120 PDR. Therefore, going from this border deeper into the condensation, we expect to find regions with different physical conditions and chemical abundances. The richest variety of molecules is found in the direction of YSOs. Particularly, we considered 5 intermediate and high-mass YSOs found by \citet{Figueira_2017}. In the selected spectral range and having a threshold of 3~$\sigma$ we identified 38 molecules towards RCW\,120~S2, including isotopologues. In the present paper, we limit ourselves by only 20 molecules, see Table~\ref{tab:all_molecules}), which produce the brightest lines. Specific emission lines of these molecules are presented in Fig.~\ref{fig:detectedlines}. Detected molecules vary from diatomic species (e.~g. SO, CO) to complex organic molecules (COMs), such as CH$_3$OH, CH$_3$CN, CH$_3$CCH, which are usually observed in hot molecular cores. To our knowledge, this is the first sensitive molecular survey of the RCW\,120 region at 1~mm.

\begin{table*}
    
    	\caption{Detected spectral lines of the twenty detected molecules and proprties of the lines: transitions, rest frequencies, upper level energies $E_{\rm u}$. All the information was taken from the CDMS database. The integrated intensity maps for all the detected molecules along with the brightness temperature maps are displayed in Fig.~\ref{fig:integrated_intensity_map} and Fig.~\ref{fig:Tmb_map}. }
            \medskip
    	\label{tab:all_molecules}
    	\begin{tabular}{l|l|c|c} 
   	 \hline
   	 Molecule    	& Transition    	& Frequency 	& E$_\text{up}$  \\ 
   	             	&               	& (MHz)     	& (K)        	\\ 
   	 \hline
	HDCO     & J$_{K_\text{a},K_\text{c}}$=3$_{1, 2}$--2$_{1, 1}$    &    201341.362    & 27.3   \\ 
	$^{\text{a}}$CH$_{3}$CCH     & J$_{K}$=12$_{0}$--11$_{0}$    &    205080.732    & 64.0   \\ 
	$^{\text{a}}$CH$_{3}$CN    & J$_K$=11$_{0}$--10$_{0}$    &    202355.510    & 58.3    \\ 
	DCO$^+$    & J=3--2    &    216112.582    & 20.7    \\ 
	H$_2$S    & J$_{K_\text{a},K_\text{c}}$=2$_{2, 0}$--2$_{1, 1}$    &    216710.437    & 84.0    \\ 
	SiO    &  J=5--4    &    217104.919    & 31.3    \\ 
	DCN    & J=3--2    &    217238.538    & 20.9   \\ 
	c-C$_{3}$H$_{2}$ 	 & J$_{K_\text{a},K_\text{c}}$=6$_{0, 6}$--5$_{1, 5}$    &    217822.148    & 38.6    \\ 
	H$_{2}$CO    & J$_{K_\text{a},K_\text{c}}$=3$_{0, 3}$--2$_{0, 2}$    &    218222.192    & 21.0   \\ 
	C$^{18}$O    & J=2--1    &    219560.354    & 15.8    \\ 
	HNCO    & J$_{K_\text{a},K_\text{c}}$=10$_{0, 10}$--9$_{0, 9}$    &    219798.274    & 58.0    \\ 
	$^{13}$CO    & J=2--1    &    220398.667    & 15.9    \\ 
	$^{\text{a}}$CH$_{3}$CN    & J$_K$=12$_{0}$--11$_{0}$    &    220747.261    & 68.9    \\ 

	 $^{\text{a}}$CH$_{3}$CCH     & J$_{K}$=13$_{0}$--12$_{0}$    &    222166.971    & 74.6   \\ 
    
	 $^{\text{a}}$CH$_{3}$CN    & J$_K$=13$_{0}$--12$_{0}$    &    239137.917    & 80.3   \\ 
	 $^{\text{a}}$CH$_{3}$CCH     & J$_{K}$=14$_{0}$--13$_{0}$    &    239252.294    & 86.1   \\ 
	H$_2$CS    & J$_{K_\text{a},K_\text{c}}$=7$_{0, 7}$--6$_{0, 6}$    &    240266.872    & 46.1    \\ 
	C$^{34}$S    &   J=5--4    &    241016.089    & 27.8    \\ 
	$^{\text{a,b}}$CH$_{3}$OH, vt=0-2    & J$_{K_\text{a},K_\text{c}}$=5$_{0, 5}$--4$_{0, 4}$, A$^{+}$    &    241791.352    & 34.8 \\ 
	CS    &   J=5--4    &    244935.557    & 35.3    \\ 
	SO    & N$_\text{J}$=6$_{6}$--5$_{5}$ 		 &    258255.826    & 56.5    \\ 
	H$^{13}$CN    & J=3--2 		 &    259011.798    & 24.9   \\ 
	H$^{13}$CO$^+$    &  J=3--2    &    260255.339    & 25.0   \\ 
	 SiO    &  J=6--5    &    260518.009    & 43.8    \\ 
	$^{\text{a}}$CCH    & N$_\text{J}$=3$_{7/2}$--2$_{5/2}$  F=4--3 & 262004.260   	 & 25.1   \\ 
   	 \hline
	\multicolumn{4}{l}{$^{\text{a}}$ Properties of the brightest component of the detected line series}\\
	\multicolumn{4}{l}{$^{\text{b}}$ Other lines of methanol are shown in Table~\ref{tab:RD_CH3CN_param}.}\\
    \end{tabular}
\end{table*}

\subsection{Maps of molecular emission}

\begin{figure*}
    	\includegraphics[width=1.95\columnwidth]{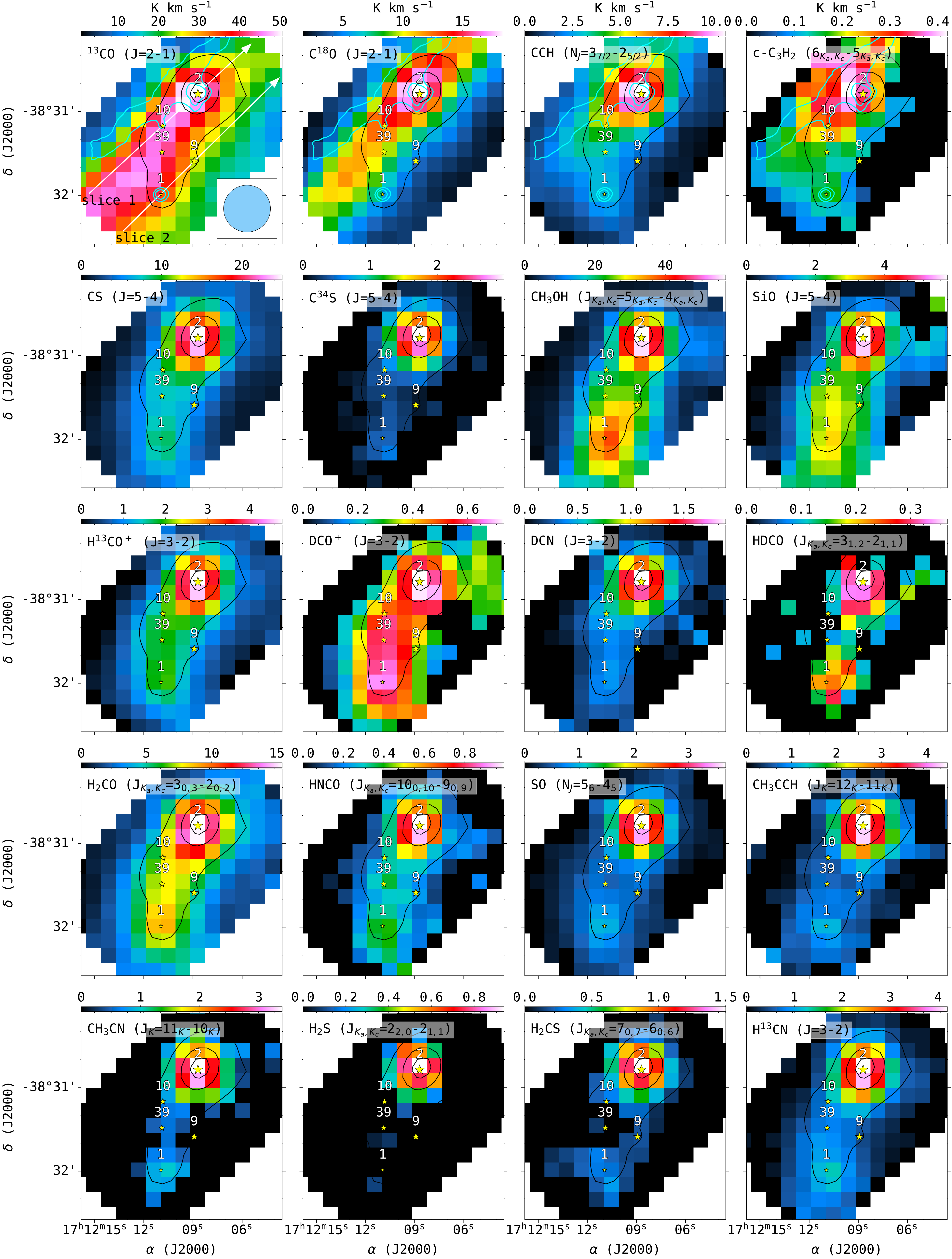}
    	\caption{Integrated intensity maps (0-th moment maps) of selected emission lines of the twenty detected molecules. The YSOs are marked by yellow stars, the sizes of the symbols are proportional to their masses. The black contours show the emission of dust at 870~$\mu$m, the contour levels are 2.0, 6.0, 10.0 Jy/beam. The blue contours show the emission of dust at 70~$\mu$m, the contour levels are 0.47, 1.0 Jy/pixel. The white arrows, labelled as `slice 1' and `slice 2', indicate the chosen directions for plotting position-velocity~(PV) diagrams, discussed in Sec.~\ref{sec:PV_diag}. For CH$_{3}$CN and CH$_{3}$CCH maps we considered K = 0,1,2,3. For CH$_{3}$OH map we considered methanol emission lines within 241680--241910~MHz frequency range (see Table~\ref{tab:RD_CH3CN_param}). The blue circle in the bottom-right corner of the $^{13}$CO panel represents the degraded APEX beam size.}
    	\label{fig:integrated_intensity_map}
\end{figure*}

\begin{figure*}
    \includegraphics[width=1.95\columnwidth]{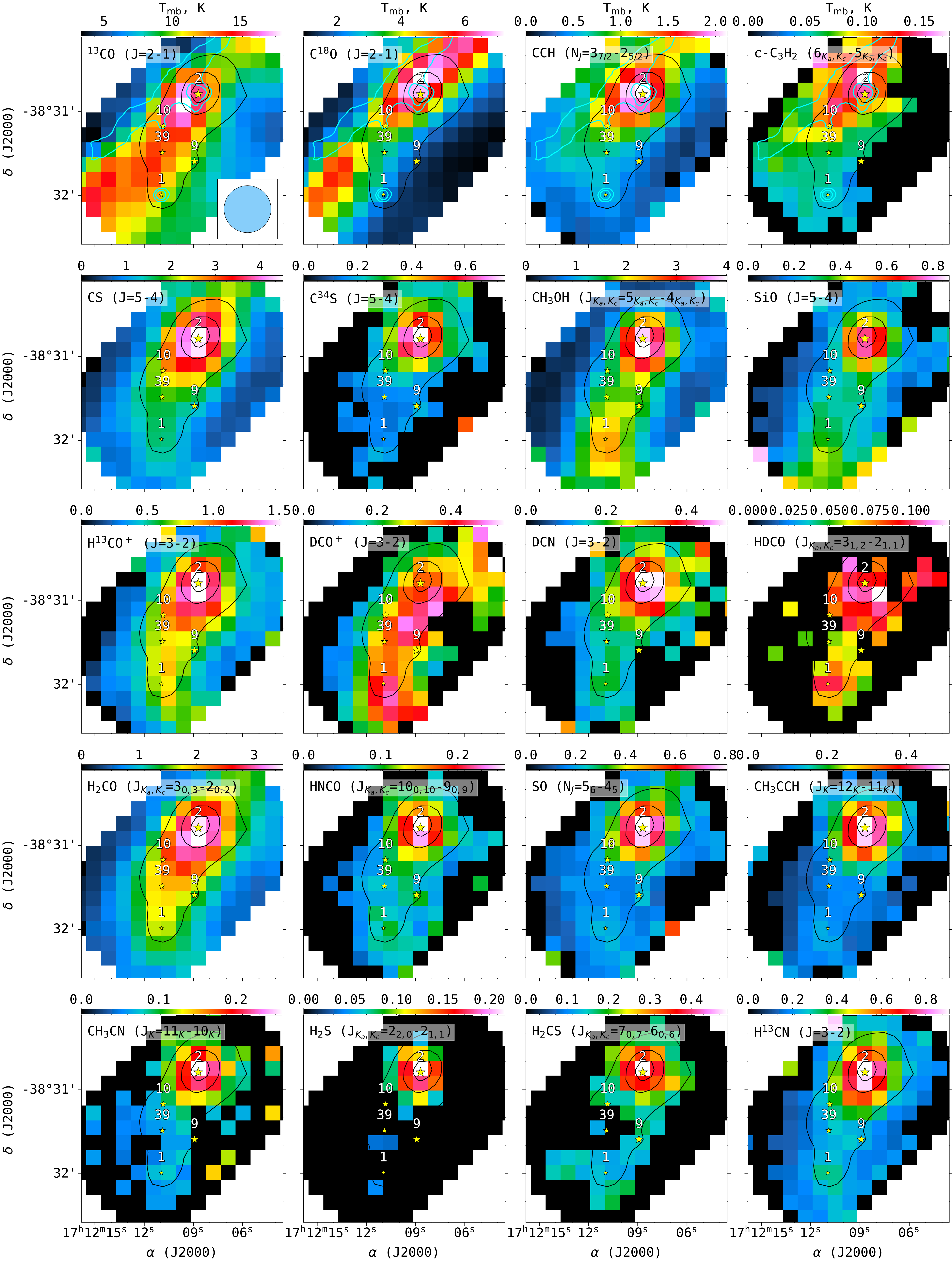}
    \caption{The peak intensity maps of selected emission lines of the twenty detected molecules. YSOs are marked by yellow stars, the sizes of the symbols are proportional to the mass of each YSO. The black contours show the emission of dust at 870~$\mu$m, the contour levels are 2.0, 6.0, 10.0 Jy/beam. The blue contours show the emission of dust at 70~$\mu$m, the contour levels are 0.47, 1.0 Jy/pixel. For CH$_{3}$CN and CH$_{3}$CCH maps we considered K = 0,1,2,3. For CH$_{3}$OH map we considered methanol emission lines within 241680--241910~MHz frequency range (see Table~\ref{tab:RD_CH3CN_param}). The blue circle in the bottom-right corner of the $^{13}$CO panel represents the degraded APEX beam size. }
\label{fig:Tmb_map}
\end{figure*}

In order to investigate spatial distribution of molecular emission around the YSOs and on the border of the RCW\,120~PDR, we selected the brightest detected molecular lines and plotted integrated intensity maps shown in Fig.~\ref{fig:integrated_intensity_map}. Maps of the peak intensities for the same lines are shown in Fig.~\ref{fig:Tmb_map}. A threshold of $3\sigma$ above the noise level was employed, and the resolution of all the maps was degraded to 33.6$^{\prime \prime}$ in order to make the signal-to-noise level higher and prepare the maps for easy comparison. Based on visual inspection of the molecular emission maps, we classified them into three groups.

The first group includes species, which have extended emission from northeast to southwest of the molecular condensation parallel to the RCW\,120 PDR. The group consists of four molecules: $^{13}$CO, C$^{18}$O, CCH and c-C$_3$H$_2$. These molecules are ubiquitous in PDR, and their emission is seen along the dust emission at 70 $\mu$m, represented with cyan contours in Fig.~\ref{fig:integrated_intensity_map}. While emission of C$^{18}$O is distributed from southeast to northwest with a peak intensity towards S2, bright emission of $^{13}$CO is more wide-spread and even protrudes towards S1. Integrated intensity maps of the CCH and c-C$_3$H$_2$ emission represent intermediate emission distribution between the first and the second groups (see below) of the integrated intensity maps. Their emission also shows an elongation in the northeast-southwest direction, but with a peak emission towards S2 and prominent extent towards S1. The molecule c-C$_3$H$_2$ has different excitation conditions because of its critical density, which is two orders of magnitude higher than for $^{13}$CO and C$^{18}$O. Nevertheless, the emission of c-C$_3$H$_2$ is also visible parallel to the RCW\,120 PDR.

The map of the C$^{18}$O~(2--1) main beam temperature, shown in Fig.~\ref{fig:Tmb_map}, reveals a dip towards S10 and S39. This dip is caused by a self-absorption effect. The main beam temperature maps for CCH and c-C$_3$H$_2$ molecules are quite similar to their integrated intensity maps, besides relatively bright emission to the northwest, which is seen in $T_{\text{mb}}$ map and hardly discernible in the integrated intensity map. This is because the CCH and c-C$_3$H$_2$ lines become bright and narrow at the northwest. We suggest that the gas density is higher at the northwest because these two hydrides have almost two orders of magnitude higher $n_{\rm crit}$ values than the CO isotopologues. While $^{13}$CO and C$^{18}$O are observed along all the PDR, the hydrides are observed towards its densest part only.

In the second group of maps, a spatial correlation is observed between the molecular emission and the 870~$\mu$m dust emission. Specifically, S1 and S2 exhibit prominent peak intensities, while the intermediate region displays relatively weaker emission levels. The second group consists of the following molecules: CS, CH$_3$OH, SiO, DCO$^+$, H$^{13}$CO$^+$, H$_2$CO, HNCO, HDCO. 
The molecular emission lines from this group display similar distributions with a compact peak emission towards S1 and S2 and a 1.5--3 times fainter homogeneous emission bridge between them. We additionally note that methanol and silicone oxide also have a bright emission feature to the south of the molecular condensation under the study.

In the peak main beam temperature map, the DCO$^+$ molecular emission looks as several separated bright spots of comparable brightness towards all the YSOs under the study, while the integrated intensity map reveals more continuous emission along the contours of the 870~$\mu$m dust continuum emission. There is also a shift between the brightest area of the integrated and the peak emission of DCO$^+$ molecule. The latter is shifted farther from the PDR. 

The third group of maps predominantly shows molecular emission emanating from the S2 YSO and barely notable in the vicinity of S1.  This group comprises molecules such as C$^{34}$S, SO, DCN, CH$_3$CCH, CH$_3$CN, H$_2$S, H$_2$CS, H$^{13}$CN. Integrated intensity of the nitrogen-bearing compounds (DCN, CH$_3$CN, H$^{13}$CN) is symmetric around S2. Meanwhile the asymmetric distribution of molecular emission towards S2 appears for sulphur-bearing species (C$^{34}$S, SO, H$_2$S, H$_2$CS). The bright emission is shifted to the east relative to S2. Distribution of CH$_3$CCH emission has the highest integrated intensity towards S2 and unlike other maps in this group, its emission is somewhat symmetric.

\subsection{Physical conditions in molecular gas}

\begin{figure*}
	\centering
	\includegraphics[ width = 2\columnwidth]{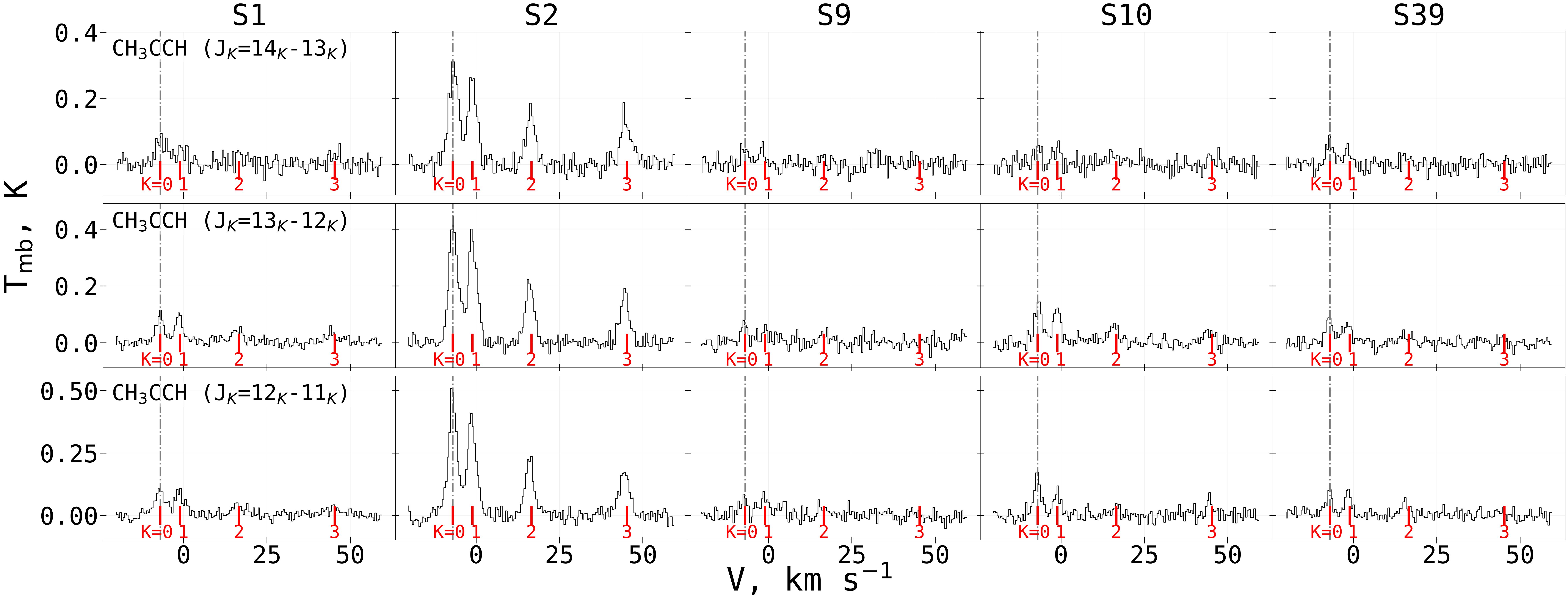}\\
	\includegraphics[ width = 2\columnwidth]{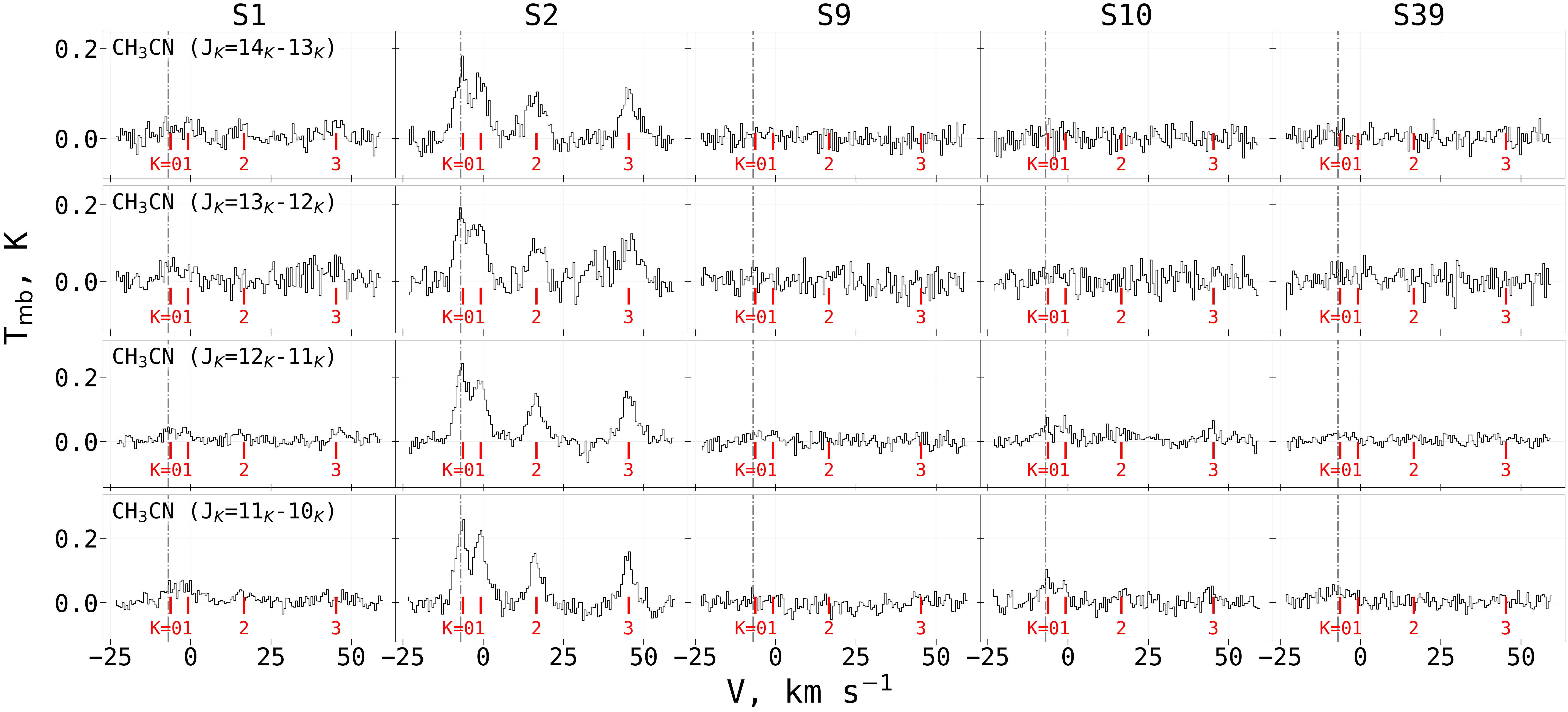}\\
	\caption{Spectral lines, which are used for LTE analysis, towards five sources. The dashed grey line marks $V_\text{LSR}$ at --7~km~s$^{-1}$.  Upper panel: CH$_3$CCH K-ladder lines, up to bottom: J$_{K}$=14$_{K}$--13$_{K}$, J$_{K}$=13$_{K}$--12$_{K}$, J$_{K}$=12$_{K}$--11$_{K}$, . Lower panel: CH$_3$CN K-ladder lines, up to bottom: J$_{K}$=14$_{K}$--13$_{K}$, J$_{K}$=13$_{K}$--12$_{K}$, J$_{K}$=12$_{K}$--11$_{K}$, J$_{K}$=11$_{K}$--10$_{K}$}
	\label{fig:cores_RD}
\end{figure*}

In the observed frequency range, we detected $J_{K}$=14$_{K}$--13$_{K}$, $J_{K}$=13$_{K}$--12$_{K}$, $J_{K}$=12$_{K}$--11$_{K}$ line series for CH$_3$CCH and  $J_{K}$=14$_{K}$--13$_{K}$, $J_{K}$=13$_{K}$--12$_{K}$, $J_{K}$=12$_{K}$--11$_{K}$, $J_{K}$=11$_{K}$--10$_{K}$ for CH$_3$CN. For these transitions, we applied the rotational diagram technique described in Sec.~\ref{sec:rotation_diagrams} to determine physical conditions in the dense molecular condensation with particular attention to the YSOs. We also use the four brightest lines of the methanol $J_{K_\text{a},K_\text{c}}$=5$_{K_\text{a},K_\text{c}}$--4$_{K_\text{a},K_\text{c}}$ series to estimate column density of this molecule through the whole obtained map. Integrated intensities of the used emission lines towards the YSOs are given in Table~\ref{tab:RD_CH3CN_param}. The spectra of the CH$_3$CCH and CH$_3$CN line series in the YSOs are shown in Fig.~\ref{fig:cores_RD} with the spectral resolution, degraded to 0.6\kms{} to increase signal-to-noise ratio. 

\begin{figure}
	\centering
	\includegraphics[ width = \columnwidth]{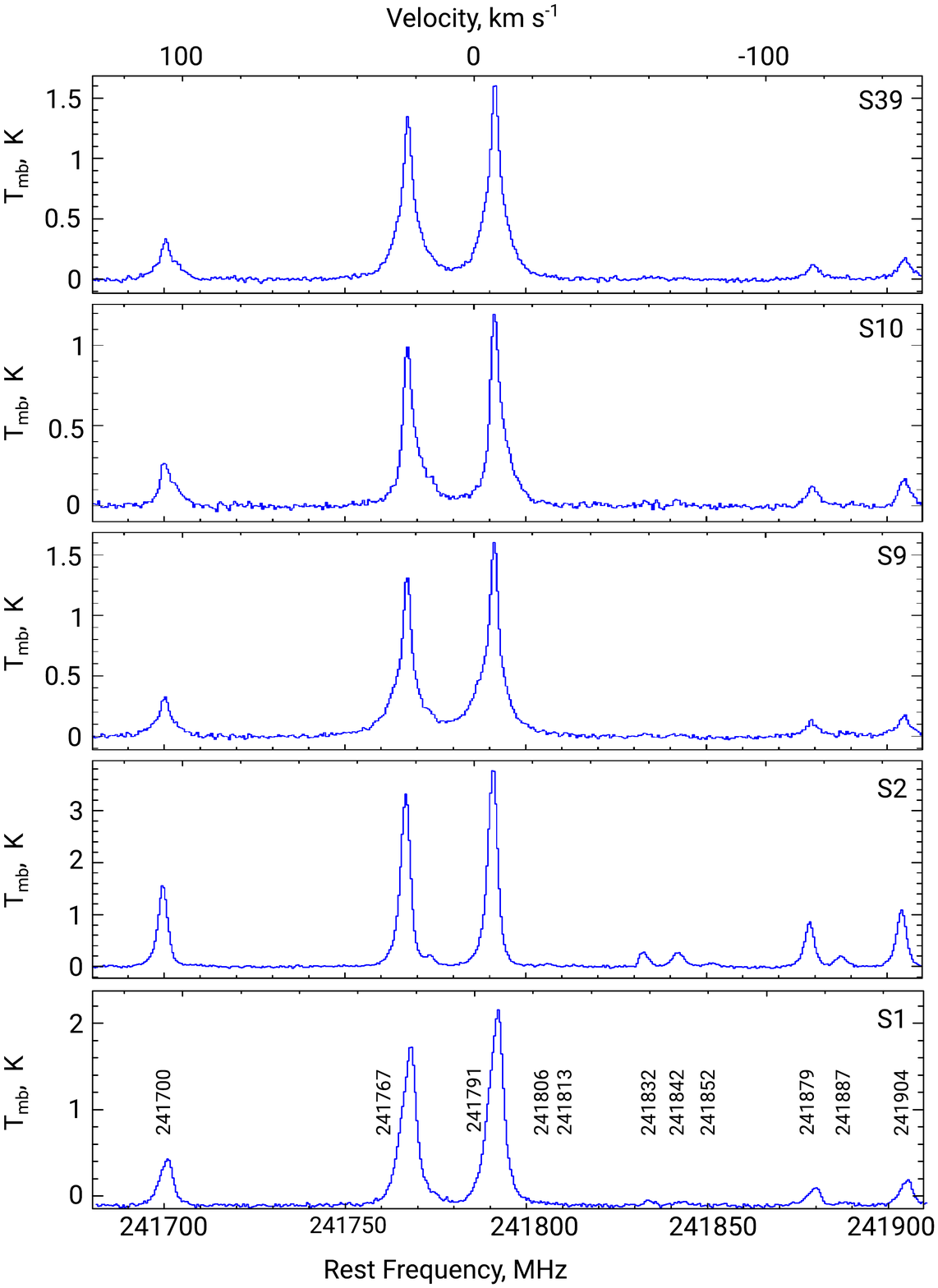}\\
	\caption{ Methanol lines in the range 241680-241910 MHz towards five YSOs. Line frequencies in MHz are given on the bottom panel. The names of YSOs are shown at upper right corners.}
	\label{fig:methanol_spectra}
\end{figure}

\subsubsection{CH$_3$CCH}

\begin{figure}
	\centering
	\includegraphics[width = \columnwidth]{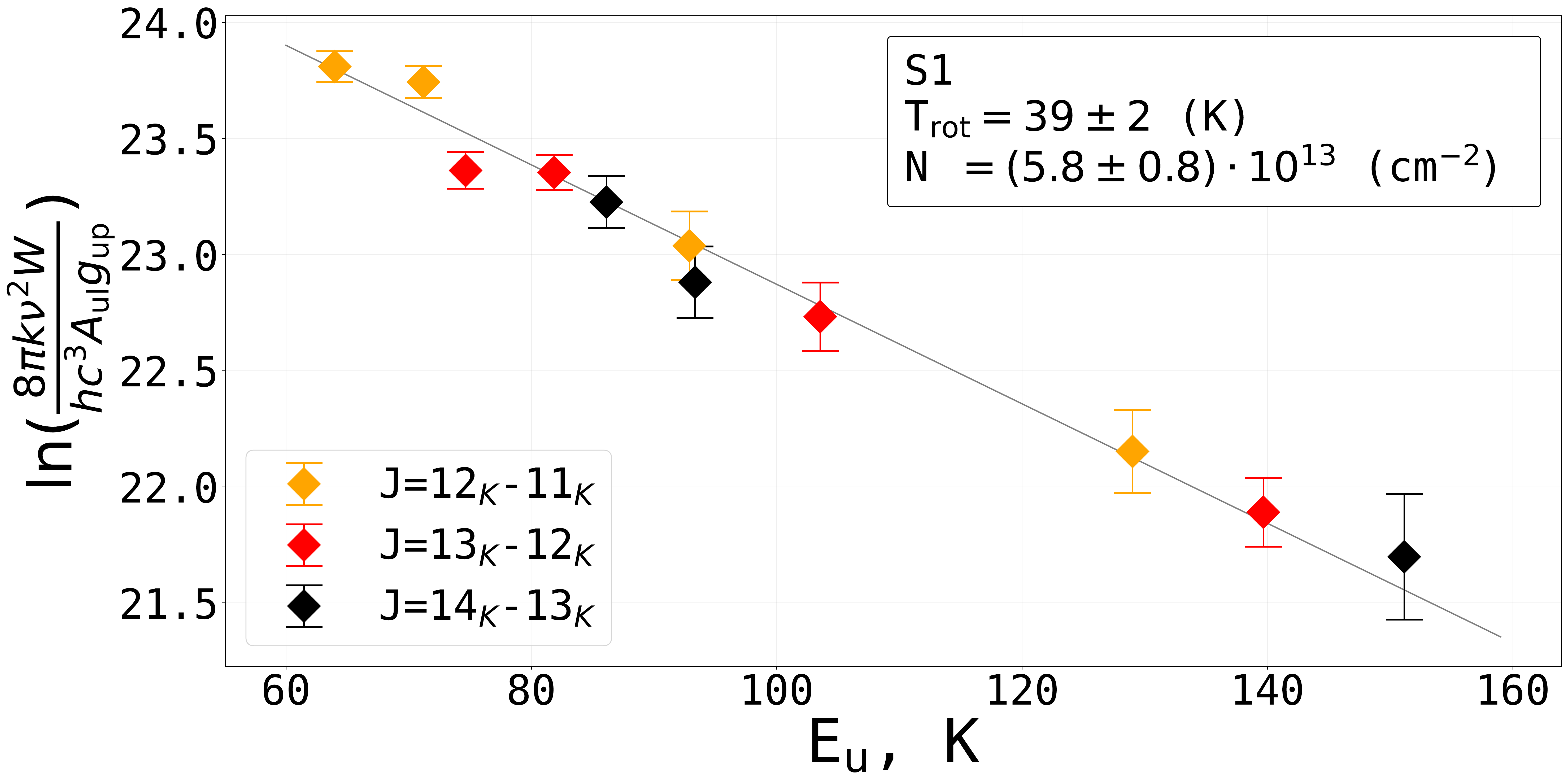}\\
	\includegraphics[width = \columnwidth]{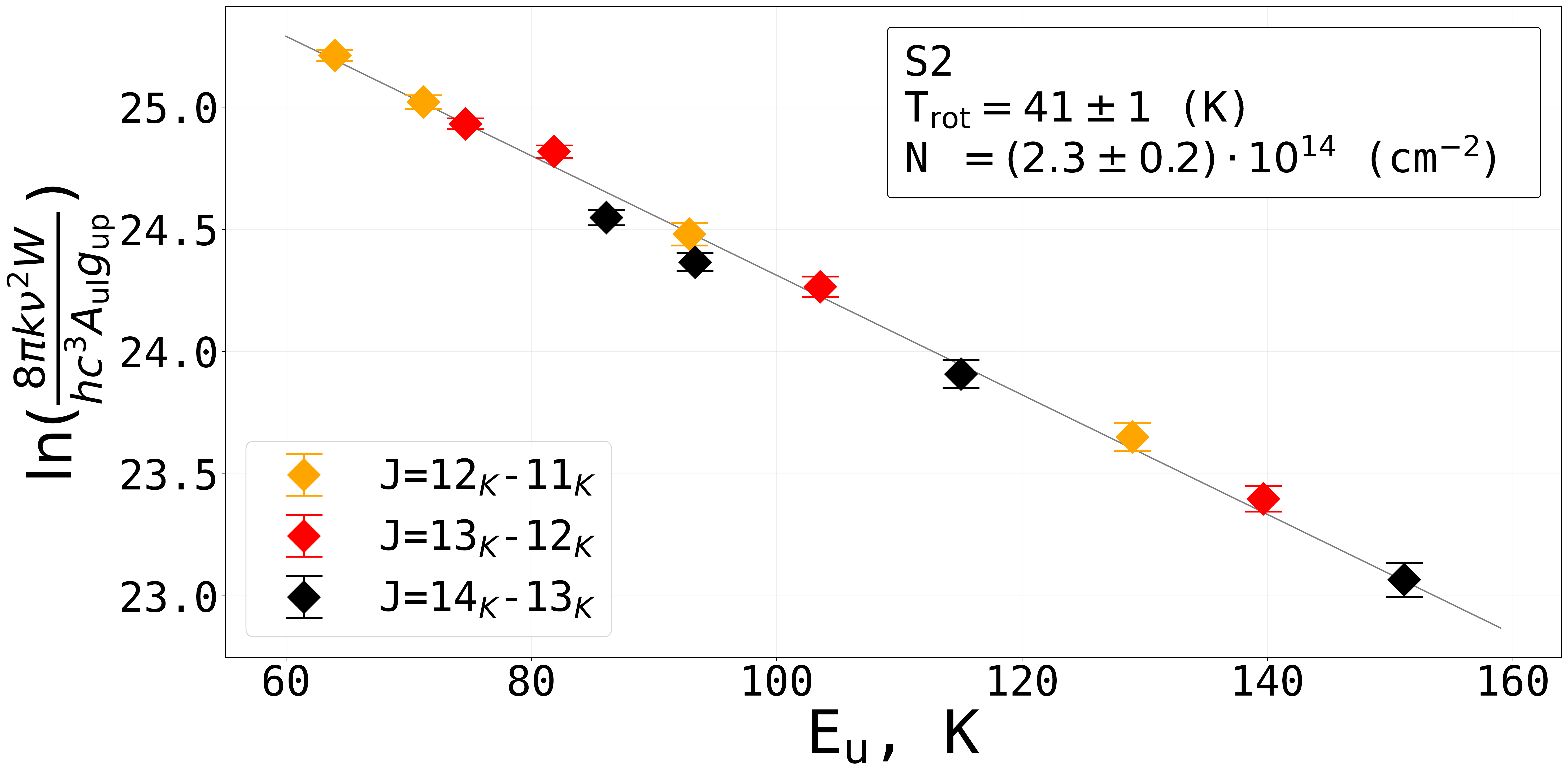}\\
	\includegraphics[width = \columnwidth]{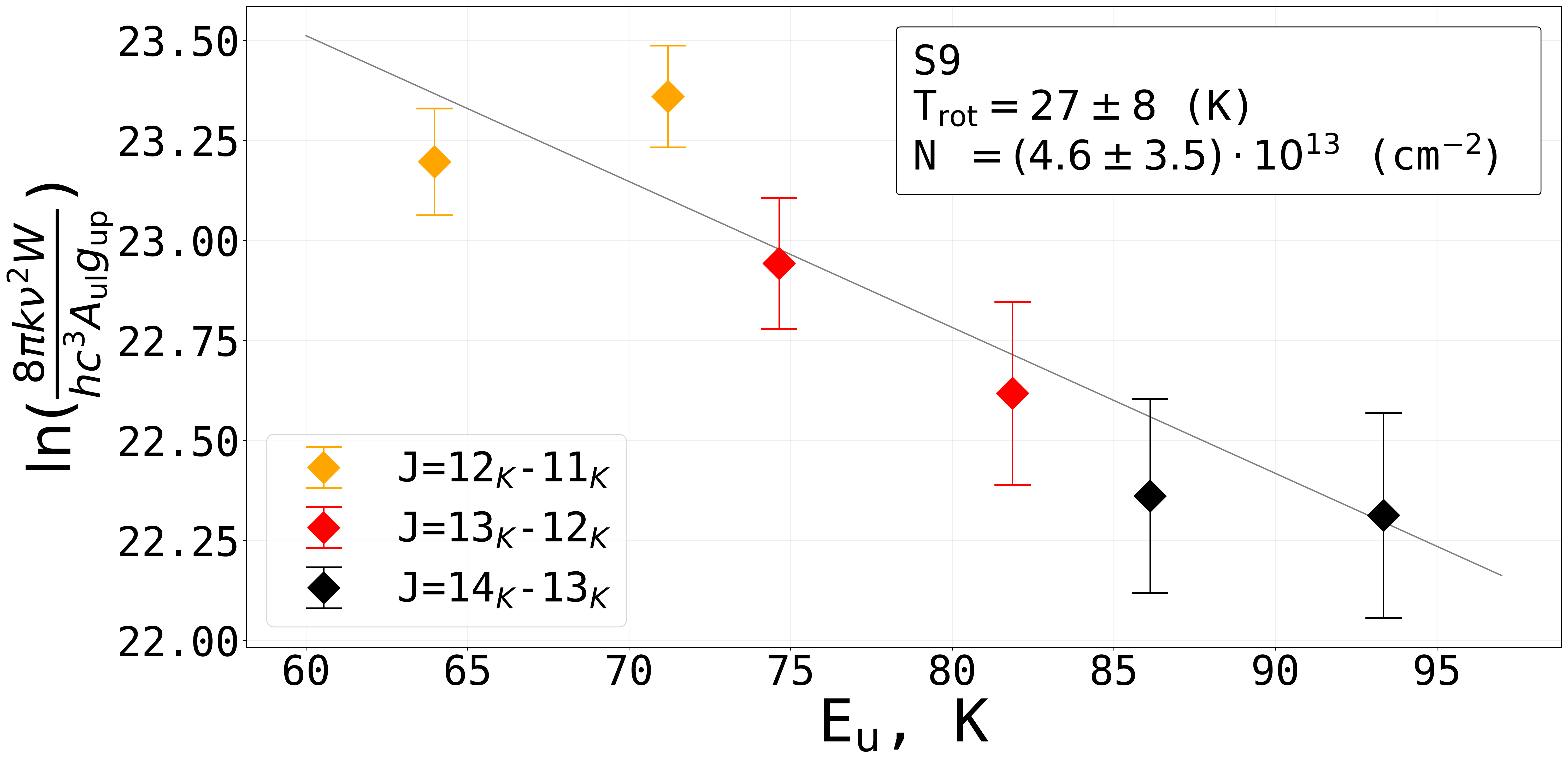}\\
	\includegraphics[width = \columnwidth]{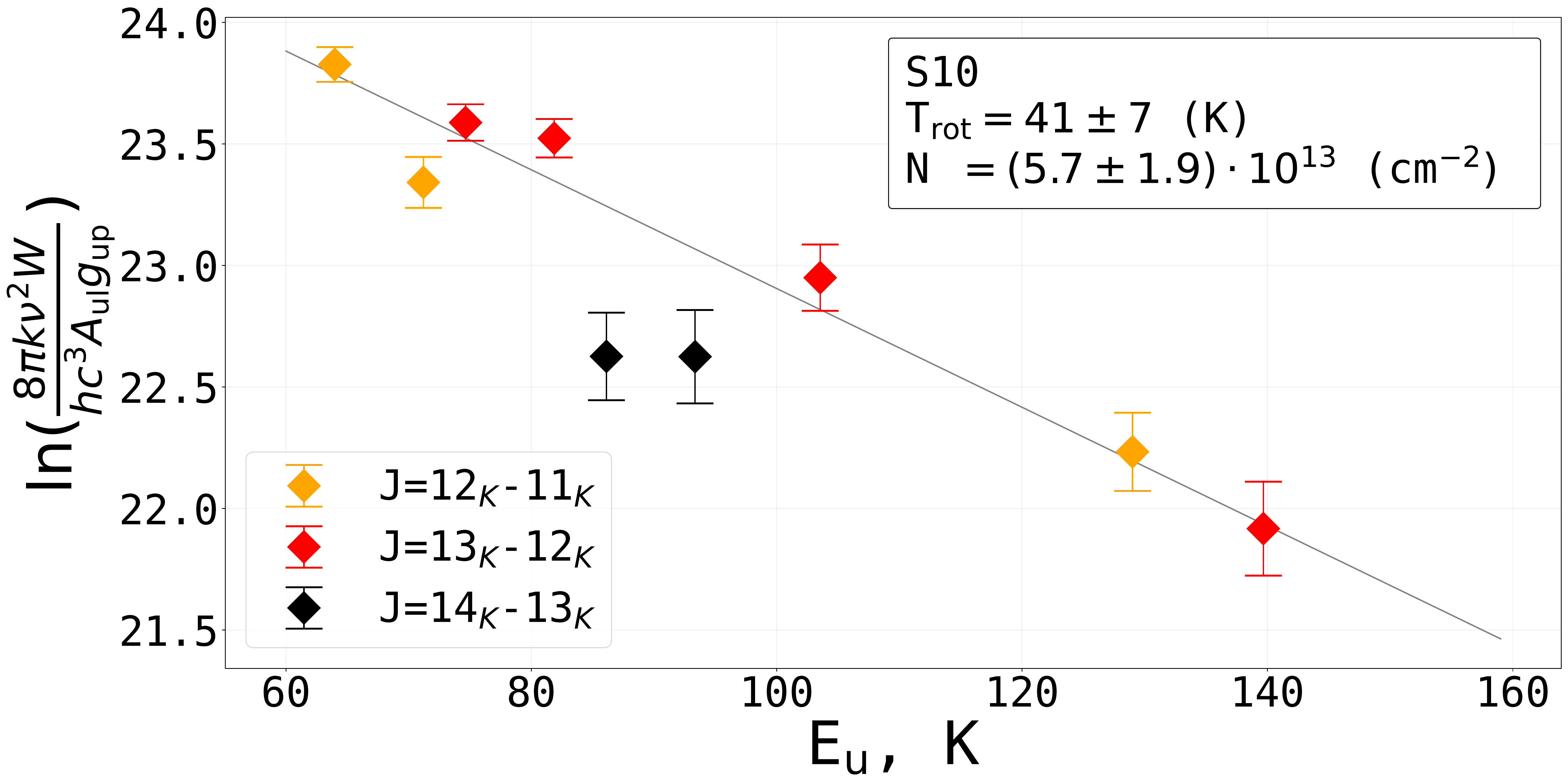}\\
	\includegraphics[width = \columnwidth]{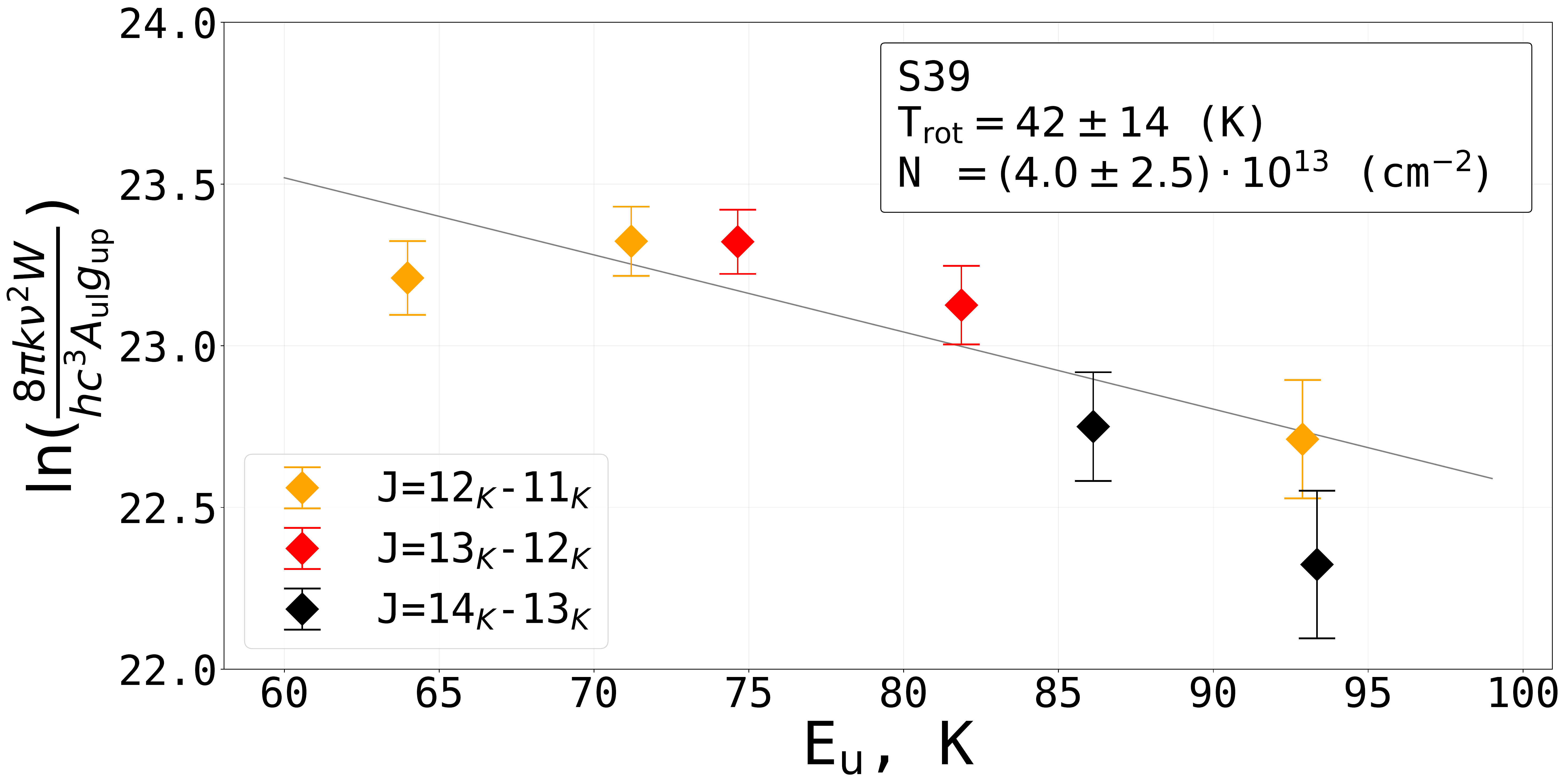}\\
	\caption{Rotational diagrams of the studied CH$_3$CCH line series. We show only transitions for S/N ratio $> 3$.}
	\label{fig:cores_RD_ch3cch}
\end{figure}

The rotational diagrams of CH$_3$CCH line series in YSOs are shown in Fig.~\ref{fig:cores_RD_ch3cch}. Given that our data in rotational diagrams of S1, S2 show overall linear trend we can suggest that our LTE assumption is reasonable. We obtained rotational temperatures of CH$_3$CCH and its column density in all the considered YSOs, see results in Table~\ref{tab:RD_CH3CCH_param}. The $K=2$ and $K=3$ components of the line series towards S9 and S39 are relatively weak. As a result, we only used the $K=0$ and $K=1$ for these YSOs (and $K=3$ in the case of the lowest observed $J$ transition for S39). This limitation led to increased uncertainties in the analysis of these YSOs compared to others where a larger number of transitions was considered. We find from Table~\ref{tab:RD_CH3CCH_param} that the $T_{\rm rot}$ values lie in the range of 30--40~K in the YSOs. The values of $N_{\rm CH_3CCH}$ are similar in all the YSOs except of S2, where the column density is about $4-5$ times higher. In order to calculate relative abundance of CH$_3$CCH, we divided $N_{\rm CH_3CCH}$ by a column density of hydrogen atoms $N_{\rm HI+H_2}$ (shown in Fig.~\ref{fig:farIRDWiebe}) found from the analysis of the dust far-IR emission.

\begin{table*}
\caption{Rotational temperatures, column densities and abundances relative to hydrogen nuclei for CH$_3$CCH, CH$_3$CN and SiO towards observed hot molecular cores in RCW\,120.}
\label{tab:RD_CH3CCH_param}
\begin{tabular}{clccclccclccccc} 
\hline
YSO	&  \multicolumn{3}{c}{CH$_3$CCH} & &  \multicolumn{3}{c}{CH$_3$CN} & &  \multicolumn{3}{c}{SiO} & & T$_{\text{dust}}$ & N(HI+H$_{2}$)\\
                	\cline{2-4} \cline{6-8} \cline{10-12} 
    	& T$_{\text{rot}}$   & N$_{\text{tot}}$   &  $X$  &   & T$_{\text{rot}}$ & N$_{\text{tot}}$&  $X$ &   & T$_{\text{rot}}$ & N$_{\text{tot}}$&  $X$ & &   \\
      &   	(K)      	& (10$^{13}$~cm$^{-2}$)    	&  (10$^{-9}$)  &   &   	(K)    	& (10$^{12}$~cm$^{-2}$) & (10$^{-10}$)  &   &   	(K)    	& (10$^{12}$~cm$^{-2}$) 	&  	(10$^{-11}$)    & &  (K) &  (10$^{22}$~cm$^{-2}$)	\\
\hline     	 
1   	&  	39$\pm$ 2	&  5.8$\pm$0.8 & 5.2  &   & 	61$\pm$ 16   & 1.5$\pm$0.6  & 1.4 &   & 12.1$\pm$ 0.4   & 8.5$\pm$0.3  & 12.2 & & 22.20& 1.11	\\
2   	&  	41$\pm$ 1	&  23$\pm$2  &   6.2 &  & 	58$\pm$ 4	& 6.5$\pm$0.7  & 1.8 &   & 18.5$\pm$ 0.7	& 9.9$\pm$0.3  & 4.10 & & 21.49& 3.71	\\
9   	&  	27$\pm$ 8	&  4.6$\pm$3.5 & 4.7  &   & 	$\cdots$ 	& $\cdots$       	&  &   & 14.1 $\pm$ 0.7   & 5.2$\pm$0.3  & 8.99  & & 21.74& 0.98	\\
10  	&  	41$\pm$ 7	&  5.7$\pm$1.9 & 4.0  &   & 	52$\pm$ 11   & 1.7$\pm$0.5  & 1.2 &   & 14.2 $\pm$ 0.1   & 4.4$\pm$0.1  & 4.71 & & 23.04 & 1.42	\\
39  	&  	42$\pm$ 14   &  4.0$\pm$2.5 & 3.2  &   &  	$\cdots$ 	& $\cdots$       	& &   & 16.3 $\pm$ 1.3  & 4.9$\pm$0.9   & 6.19 & & 22.02 & 1.26	\\
\hline       	 
\end{tabular}
\end{table*}

\subsubsection{CH$_3$CN}

\begin{figure} [h]
	\centering
	\includegraphics[ width = \columnwidth]{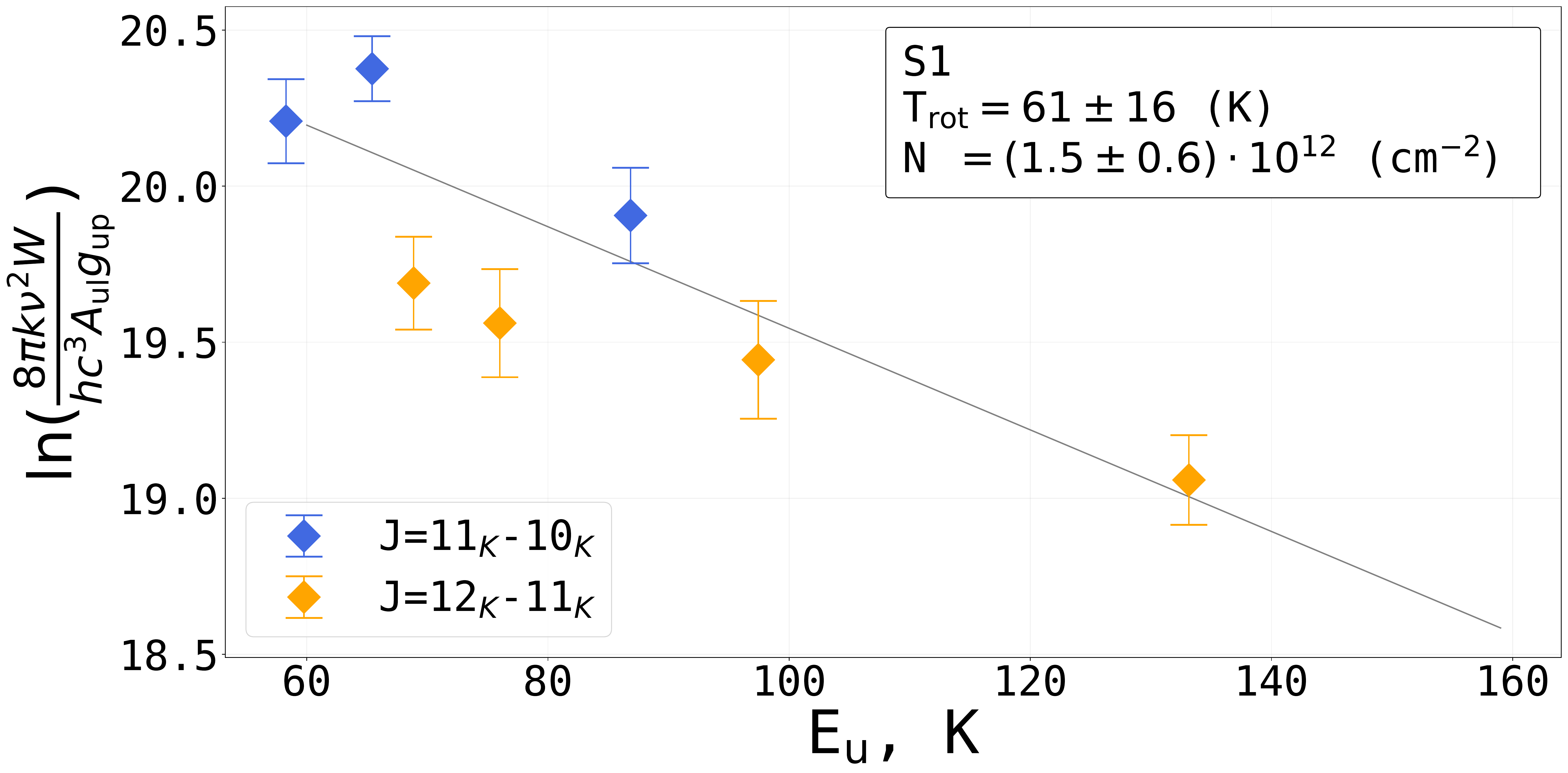}\\
	\includegraphics[ width = \columnwidth]{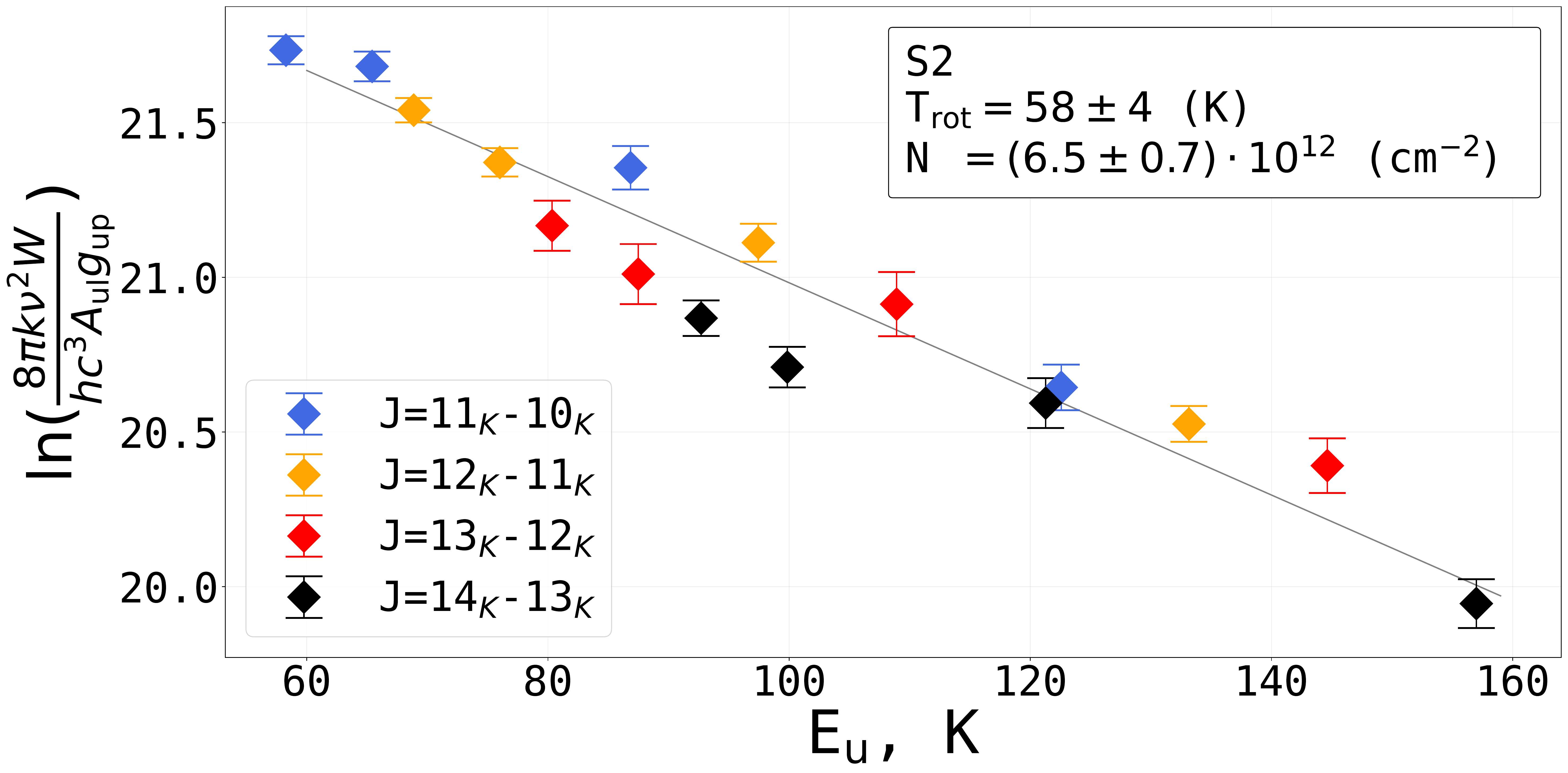}\\
	\includegraphics[ width = \columnwidth]{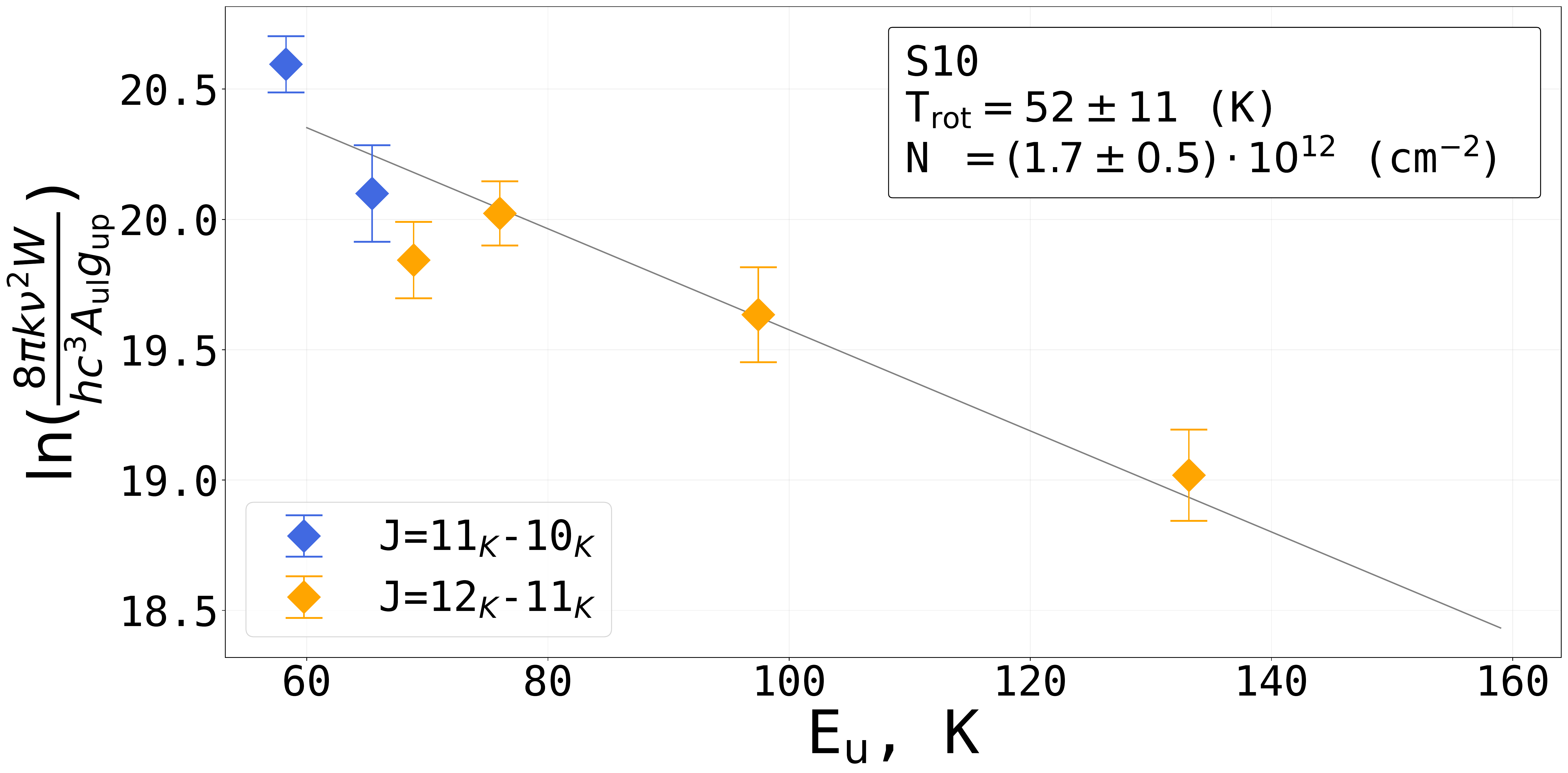}\\
	\caption{Rotational diagrams of the studied CH$_3$CN line series. We show only transitions for S/N ratio > 3.}
	\label{fig:cores_RD_ch3cn}
\end{figure}

For CH$_3$CN, we adopted the same data reduction and a similar approach by selecting four lines of each transition and presenting them in the lower panel of Fig.~\ref{fig:cores_RD}. The signal-to-noise ratio for observing the CH$_3$CN line series towards S9 and S39 is insufficient to obtain reliable results. In contrast, S2 was the only YSO where we successfully detected all four lines for each line series. The rotational diagrams are shown in Fig.~\ref{fig:cores_RD_ch3cn} and physical parameters are presented in Table~\ref{tab:RD_CH3CCH_param}. We find that $T_{\rm rot}$ of the CH$_3$CN lines are higher $\approx 20$~K for S1 and S2 and $\approx 10$~K for S10 than for the CH$_3$CCH lines. Therefore,  CH$_3$CN  emission traces warmer gas than CH$_3$CCH emission, forming an onion-like structure. Together with the compact spatial distribution of CH$_3$CN, we suggest that its emission traces closer vicinity of the YSOs with higher density and temperature than CH$_3$CCH. Column density $N{\rm (CH_3CN)}$ towards S2 is $\approx 4$ times higher than in S1 and S10. This is the same contrast between the YSOs as we observe for CH$_3$CCH. Therefore, we suggest that similar factors or processes control formation of these molecules towards the YSOs.

\subsubsection{SiO}

Rotational temperature of SiO is $2-3$ times less compared with CH$_3$CCH and lies within the interval 12--19~K. The SiO lines used for the diagrams have $E_{\rm u}$ values from 6 to 40~K. These values are a factor of few less than for CH$_3$CCH and especially for CH$_3$CN molecules. Therefore, with the SiO lines we observe less excited molecular gas. Column densities $N{\rm (SiO)}$ in S1 and S2 are almost the same and exceed the values towards all other YSOs only by a factor of 2.

\subsubsection{CH$_3$OH}\label{sec:methanolanalysis}

Methanol is the only molecule, for which we apply non-LTE analysis estimating physical parameters of the gas towards the YSOs directly. In Table~\ref{tab:RD_CH3CN_param}, we summarise 22 methanol lines used for the analysis and provide their integrated intensities. For example Fig.~\ref{fig:methanol_spectra}  methanol lines that were observed in the range 241680-241910 MHz. 
All other lines are distributed in different parts of the observed frequency range. Methanol lines from different series make the non-LTE analysis sensitive to both gas temperature and the gas density. Estimated physical conditions as well as their confidence intervals are presented in Table~\ref{tab:LVGmod_res}. Comparing the best-fit gas temperature from the methanol analysis with the $T_{\rm rot}$ values for CH$_3$CCH, we see that the former is $\approx 10$~K lower than the latter. However, both values agree with each other in the 95\% confidence intervals for methanol. The best-fit methanol column densities are around $2.5 \times 10^{15}$~cm$^{-2}$ in S1, $5.0 \times 10^{14}$~cm$^{-2}$ in S2 and approximately two times lower in S9, S10, and S39. The sensitivity of our methanol lines is higher to the gas number density and allows concluding that the density is the highest in S2. Relative methanol abundance according to our results $3\times 10^{-6}$ in S1, S2 and S9 and only $3\times 10^{-8}$ in the other YSOs. Since the ratio of the $N{\rm (CH_3OH)}$ values between the YSOs is not the same as for $N{\rm (CH_3CCH)}$ and $N{\rm (CH_3CN)}$, namely, the contrast of $N{\rm (CH_3OH)}$ is not so high over the YSOs, we propose that different processes or factors control $N{\rm (CH_3OH)}$ in the molecular condensation. As we see, the beam-filling factor is $\approx 10 \%$ at least towards the YSOs.

\begin{table*}
\centering
\caption{Estimated physical conditions in the cores that were estimated with non-LTE anaysis. 95\% confident intervals by parameters  are presented in brackets. To  obtain the confidence intervals, we applied the Bayesian approach (e.g.~\cite{Ward2003}).}  
\label{tab:LVGmod_res}
\small
\begin{tabular}{c|c|c|c|c|c|c}
\hline
YSO     	&$T_{\rm k}$  & $n{\rm (H_2)}$   	&N   & $\chi^2$   	& X  & \textit{f} \\
         	& (K)        	& ($10^5 $cm$^{-3}$)      	& ($10^{14} $ cm$^{-2}$ ) &  &  ($10^{-8} $)     	& \% \\  
\hline
1      	& $20(10-30)$  	&$0.3(0.2-1.0)  $	&$25.1(19.9-31.6)$  & $116.2 $     &$316(100-316) $   & $20(10-30)$\\
2      	& $40(30-50)$  	&$10.0(5.6-17.8)$	&$5.0(4.0-6.3) $	& $373.5 $     &$316(100-316)$   & $20(10-30)$\\   
9      	& $30(20-50)$  	&$1.8(0.6-3.2)  $	&$3.0(0.2-3.8) $	& $44.9 $     &$316(10-316)$   & $10(10-20)$\\   
10     	& $30(20-40)$  	&$1.8(0.6-5.6)  $	&$1.9(1.2-3.8)  $	& $45.6 $     &$3.16(0.63-316.23)$   & $10(10-20)$\\   
39     	& $30(20-40)$  	&$1.0(0.1-3.2) $	&$2.4(1.2-3.8)  $	& $32.1 $     &$3.16(1.00-19.95)$   & $10(10-20)$\\   
\hline
\end{tabular}
\end{table*}

We apply a rotational diagram method for the entire map. We checked values of $N{\rm (CH_3OH)}$ from the LTE and non-LTE analysis and found them in agreement in contrast with excitation temperatures. Therefore, we use the LTE results below to study large-scale distribution of methanol in the dense condensation. 
\vspace{19cm}

\LTcapwidth=\textwidth%
    	\begin{longtable*}{c|c|c|c|c|cc|c|c|c|c|c} 
            
            \caption{CH$_{3}$CCH and CH$_{3}$CN and CH$_{3}$OH lines data from spectra in YSOs}
    	\label{tab:RD_CH3CN_param}\\

   	 \hline
                    	&           	&          	&               	&                   	&             	& & \multicolumn{5}{c}{$\int T_{\text{mb}} dV$}  \\
   	 Species 	& Frequency 	& $J_K$    & E$_{\text{up}}$   & $\log$A$_{\text{ij}}$ & g$_{\text{up}}$ & &\multicolumn{5}{c}{(K km s$^{-1})$}	\\
        	\cline{8-12}
   	           	&  (MHz)    	&   	&	(K)        	&                   	&             	& & S1& S2& S9& S10& S39         	\\

   	 \hline
	CH$_3$CCH       	& 205045.500 & 12$_{3}$--11$_{3}$  & 129.0 & -4.557 & 100 &  & $0.14\pm0.03$ &  $0.63\pm0.04$ & $\cdots$ & $0.15\pm0.02$  & $\cdots$  \\
   	           	& 205065.070 & 12$_{2}$--11$_{2}$  & 92.88 & -4.541 & 50  &  & $0.18\pm0.03$ &  $0.75\pm0.03$ & $\cdots$ & $\cdots$ & $0.13\pm0.02$ \\
                  	& 205076.816 & 12$_{1}$--11$_{1}$  & 71.20 & -4.532 & 50  &  & $0.37\pm0.03$ &  $1.32\pm0.04$ & $0.25\pm0.03$	& $0.25\pm0.03$  & $0.24\pm0.03$  \\
   	           	& 205080.732 & 12$_{0}$--11$_{0}$  & 63.98 & -4.529 & 50  &  & $0.40\pm0.03$ &  $1.61\pm0.04$ & $0.22\pm0.03$	& $0.40\pm0.03$  & $0.22\pm0.02$  \\
                    	& 222128.815 & 13$_{3}$--12$_{3}$  & 139.7 &  -4.447  & 108  &  & $0.13\pm0.02$ &  $0.58\pm0.03$  & $\cdots$ &  $0.13\pm0.03$  & $\cdots$ \\
                    	& 222150.010 & 13$_{2}$--12$_{2}$  & 103.5 &  -4.433  & 54   &  & $0.16\pm0.02 $&  $0.72\pm0.03$  & $\cdots$ &  $0.19\pm0.03$  & $\cdots$ \\
                    	& 222162.730 & 13$_{1}$--12$_{1}$  & 81.9  &  -4.426  & 54   &  & $0.29\pm0.02$ &  $1.27\pm0.03 $ & $0.14\pm0.03$ &  $0.35\pm0.03$  & $0.23\pm0.03$ \\
                    	& 222166.971 & 13$_{0}$--12$_{0}$  & 74.6  &  -4.423  & 54   &  & $0.30\pm0.02$ &  $1.43\pm0.03$  & $0.20\pm0.03$ &  $0.37\pm0.03$  & $0.29\pm0.03$ \\
                    	& 239211.215 & 14$_{3}$--13$_{3}$  & 151.14 &  -4.346  & 116 &  & $0.13\pm0.03	$&  $0.49\pm0.03$  & $\cdots$ & $\cdots$ & $\cdots$ \\
                    	& 239234.034 & 14$_{2}$--13$_{2}$  & 115.02 &  -4.334  & 58  &  & $\cdots$ &  $0.58\pm0.03$  & $\cdots$ & $\cdots$ & $\cdots$ \\
                    	& 239247.728 & 14$_{1}$--13$_{1}$  & 93.35 &   -4.327  & 58  &  & $0.21\pm0.03$	&  $0.94\pm0.03$  & $0.12\pm0.03$  & $0.16\pm0.03$ & $0.12\pm0.03$ \\
                    	& 239252.294 & 14$_{0}$--13$_{0}$  & 86.12 &   -4.325  & 58  &  & $0.30\pm0.03$	&  $1.13\pm0.04$  & $0.13\pm0.03$  & $0.17\pm0.03$ & $0.19\pm0.03$ \\
        	\hline
	CH$_3$CN        	& 202320.443 & 11$_{3}$--10$_{3}$  &122.57 & -3.183 & 92  &  &  $\cdots$	& $0.70\pm0.05$  & $\cdots$  & $\cdots$  & $\cdots$ \\
                    	& 202339.922  & 11$_{2}$--10$_{2}$  &86.85 &  -3.164 & 46  &  &  $0.18\pm0.03$ & $0.75\pm0.05$  & $\cdots$ & $\cdots$ & $\cdots$\\
                    	& 202351.612  & 11$_{1}$--10$_{1}$  &65.42 &  -3.153 & 46  &  &  $0.29\pm0.03$ & $1.06\pm0.05$  & $\cdots$ & $0.22\pm0.04$ & $\cdots$ \\
                    	& 202355.510  & 11$_{0}$--10$_{0}$  &58.27 &  -3.149 & 46  &  &  $0.24\pm0.03$ & $1.13\pm0.05$  & $\cdots$ & $0.36\pm0.04 $ & $\cdots$ \\
                    	& 220709.017 & 12$_{3}$--11$_{3}$  &133.16 & -3.062  & 100  &  &  $0.17\pm0.02$ & $0.75\pm0.04$  & $\cdots$ & $0.17\pm0.03$  & $\cdots$\\
                    	& 220730.261  & 12$_{2}$--11$_{2}$  &97.44 &  -3.047  & 50   &  &  $0.13\pm0.02$   & $0.70\pm0.04$  & $\cdots$ & $0.16\pm0.03$  & $\cdots$\\
                    	& 220743.011  & 12$_{1}$--11$_{1}$  &76.01 &  -3.037  & 50   &  &  $0.15\pm0.03$ & $0.93\pm0.04$  & $\cdots$ & $0.24\pm0.03$  & $\cdots$ \\
                    	& 220747.262  & 12$_{0}$--11$_{0}$  &68.87 &  -3.034  & 50   &  &  $0.17\pm0.03$ & $1.11\pm0.04$  & $\cdots$ & $0.20\pm0.03$  & $\cdots$ \\
                    	& 239096.497 & 13$_{3}$--12$_{3}$  &144.63 &  -2.952  & 108  &  &  $\cdots$ & $0.78\pm0.07$  & $\cdots$ & $\cdots$  & $\cdots$ \\
                    	& 239119.505 & 13$_{2}$--12$_{2}$  &108.92 &  -2.939 & 54   &  &  $\cdots$ & $0.68\pm0.07$  & $\cdots$ & $\cdots$  & $\cdots$ \\
                    	& 239133.313  & 13$_{1}$--12$_{1}$  &87.49 &  -2.931  & 54   &  &  $\cdots$ & $0.76\pm0.07$  & $\cdots$ & $\cdots$  & $\cdots$ \\
                    	& 239137.917  & 13$_{0}$--12$_{0}$  &80.34 &  -2.929  & 54   &  &  $\cdots$ & $0.89\pm0.07$  & $\cdots$ & $\cdots$  & $\cdots$ \\
                    	& 257482.792 & 14$_{3}$--13$_{3}$  &156.99 &  -2.852  & 116 &  &  $\cdots$ & $0.58\pm0.05$  & $\cdots$ & $\cdots$ & $\cdots$ \\
                    	& 257507.562 & 14$_{2}$--13$_{2}$  &121.28 &  -2.840  & 58  &  &  $\cdots$ & $0.57\pm0.05$  & $\cdots$ & $\cdots$ & $\cdots$ \\
                    	& 257522.428  & 14$_{1}$--13$_{1}$  &99.85 &   -2.833  & 58  &  &  $\cdots$ & $0.65\pm0.04$  & $\cdots$ & $\cdots$ & $\cdots$ \\
                    	& 257527.384 & 14$_{0}$--13$_{0}$  & 92.7 &   -2.831  & 58  &  &  $\cdots$ & $0.77\pm0.04$  & $\cdots$ & $\cdots$ & $\cdots$ \\
\hline
CH$_3$OH&205791.270  &  $1_1-2_0 A^+$  	& 16.84 &  $-4.473$&  3&&$  0.56\pm0.02$	&$ 1.55\pm0.02$	&$  0.29\pm0.02$	&$	0.30\pm0.02$	&$  	0.07\pm0.01$ \\
    	&216945.521  & $5_{-1}-4_{-2} E$   & 55.87 & $-4.916$ & 11&&$  0.29\pm0.02$	&$ 0.97\pm0.03$	&$  0.12\pm0.02$	&$	0.18\pm0.02$	&$  	0.03\pm0.01$  \\
    	&218440.063  	& $4_{-2}-3_{-1} E$& 45.46 & $-4.329$&  9&&$  2.86\pm0.03$	&$ 7.11\pm0.04$	&$  1.51\pm0.04$	&$	1.38\pm0.03$	&$  	0.36\pm0.05$  \\
    	&220078.561  	& $8_0-7_{-1} E$   & 96.61 & $-4.599$& 17&&$  0.25\pm0.02$	&$ 1.05\pm0.03$	&$  0.09\pm0.02$	&$	0.15\pm0.02$	&$  	0.03\pm0.01$ \\
    	&239746.219  	& $5_1-4_1 A^+$	& 49.06 & $-4.246$& 11&&$  1.36\pm0.03$	&$ 3.74\pm0.02$	&$  0.67\pm0.02$	&$	0.60\pm0.03$	&$  	0.15\pm0.03$ \\
    	&241700.219  	& $5_{0}-4_{0} E$  & 47.93 & $-4.219$& 11&&$  2.86\pm0.04$	&$ 6.49\pm0.06$	&$  1.44\pm0.05$	&$	1.20\pm0.04$	&$  	0.34\pm0.05$ \\
    	&241767.224  	& $5_{-1}-4_{-1} E$& 40.39 & $-4.236$& 11&&$  10.15\pm0.30$   &$ 14.03\pm0.24$   &$  5.78\pm0.30$	&$	4.26\pm0.18$	&$  	1.39\pm0.20$ \\
    	&241791.431  	& $5_0-4_0 A^+$	& 34.82 & $-4.218$& 11&&$  12.37\pm0.30$   &$ 16.19\pm0.18$   &$  6.99\pm0.40$	&$	5.17\pm0.23$	&$  	1.68\pm0.25$  \\
    	&241806.508  	& $5_4-4_4 A^+A^-$ &115.16 & $-4.662$& 11&&$   \cdots$    	&$ 0.23\pm0.03$	&$	\cdots  $ 	&$	\cdots  $   	&$	\cdots  $  \\
    	&241813.257  	& $5_{-4}-4_{-4} E$&122.72 & $-4.662$& 11&&$   \cdots$    	&$ 0.10\pm0.03$	&$	\cdots  $ 	&$	\cdots  $   	&$	\cdots  $  \\
    	&241829.646  	& $5_{4}-4_{4} E$  &130.82 & $-4.660$& 11&&$   \cdots$    	&$	\cdots  $	&$	\cdots  $ 	&$	\cdots  $   	&$	\cdots  $  \\
    	&241832.910$^1$  & $5_3-4_3 A^+$	& 84.61 & $-4.413$& 11&&$  0.24\pm0.04$	&$ 1.09\pm0.15$	&$	\cdots  $ 	&$	\cdots  $   	&$	\cdots  $  \\	 
    	&241833.104$^1$  & $5_3-4_3 A^-$	& 84.61 & $-4.413$& 11&&$  0.24\pm0.04$	&$ 1.09\pm0.15$	&$	\cdots  $ 	&$	\cdots  $   	&$	\cdots  $  \\
    	&241842.324$^2$  & $5_2-4_2 A^-$	& 72.53 & $-4.291$& 11&&$  0.21\pm0.04$	&$ 1.21\pm0.02$	&$	\cdots  $ 	&$	0.12\pm0.02$	&$	\cdots  $  \\
    	&241843.646$^2$  & $5_{3}-4_{3} E$  & 82.53 & $-4.411$& 11&&$  0.21\pm0.04$	&$ 1.21\pm0.02$	&$	\cdots  $ 	&$	0.12\pm0.02$	&$	\cdots  $  \\  	 
    	&241852.352  	& $5_{-3}-4_{-3} E$& 97.53 & $-4.409$& 11&&$   \cdots$    	&$ 0.29\pm0.03$	&$	\cdots  $ 	&$	\cdots  $   	&$	\cdots  $  \\
    	&241879.073  	& $5_{1}-4_{1} E$  & 55.87 & $-4.224$& 11&&$  1.06\pm0.04$	&$ 3.52\pm0.10$	&$  0.55\pm0.03$	&$	0.52\pm0.02$	&$  	0.13\pm0.02$  \\
    	&241887.704  	& $5_2-4_2 A^+$	& 72.53 & $-4.290$& 11&&$  0.19\pm0.03$	&$ 0.88\pm0.07$	&$	\cdots  $ 	&$	\cdots$     	&$	\cdots  $   \\
    	&241904.147$^3$  & $5_{-2}-4_{-2} E$& 60.72 & $-4.293$& 11&&$  1.53\pm0.03$	&$ 4.53\pm0.12$	&$  0.76\pm0.05$	&$	0.73\pm0.02$	&$  	0.18\pm0.03$   \\
    	&241904.645$^3$  & $5_{2}-4_{2} E$  & 57.07 & $-4.298$& 11&&$  1.53\pm0.03$	&$ 4.53\pm0.12$	&$  0.76\pm0.05$	&$	0.73\pm0.02$	&$  	0.18\pm0.03$   \\    
    	&243915.788  	& $5_1-4_1 A^+$	& 49.66 & $-4.224$& 11&&$  1.88\pm0.03$	&$ 4.62\pm0.03$	&$  0.91\pm0.05$	&$	0.74\pm0.04$	&$  	0.20\pm0.04$   \\
    	&261805.675  	& $2_{-1}-1_{0} E$ & 28.01 & $-4.254$&  5&&$  1.88\pm0.03$	&$ 3.38\pm0.03$	&$  0.92\pm0.03$	&$	0.77\pm0.04$	&$  	0.24\pm0.04$ \\  
\hline

\multicolumn{12}{l}{\parbox{1.1\linewidth}{CH$_3$OH lines were fitted in program CLASS of GILDAS package (\url{http://www.iram.fr/IRAMFR/GILDAS}) by 1 Gaussian fitting with fixed FWHM = 5 km s$^{-1}$ for S1 and 4 km s$^{-1}$ for others YSOs. $V_{lsr}$ that coincident to  maximum of emission were fixed too ($-6.59, -7.14, -6.94, -6.80, -7.05$ km s$^{-1}$ for S1, S2, S9, S10 and S39, respectively)} } \\
   	 \end{longtable*}

\subsection{Hydrogen column density and dust temperature}

\begin{figure}
    \includegraphics[width=0.85\columnwidth,]{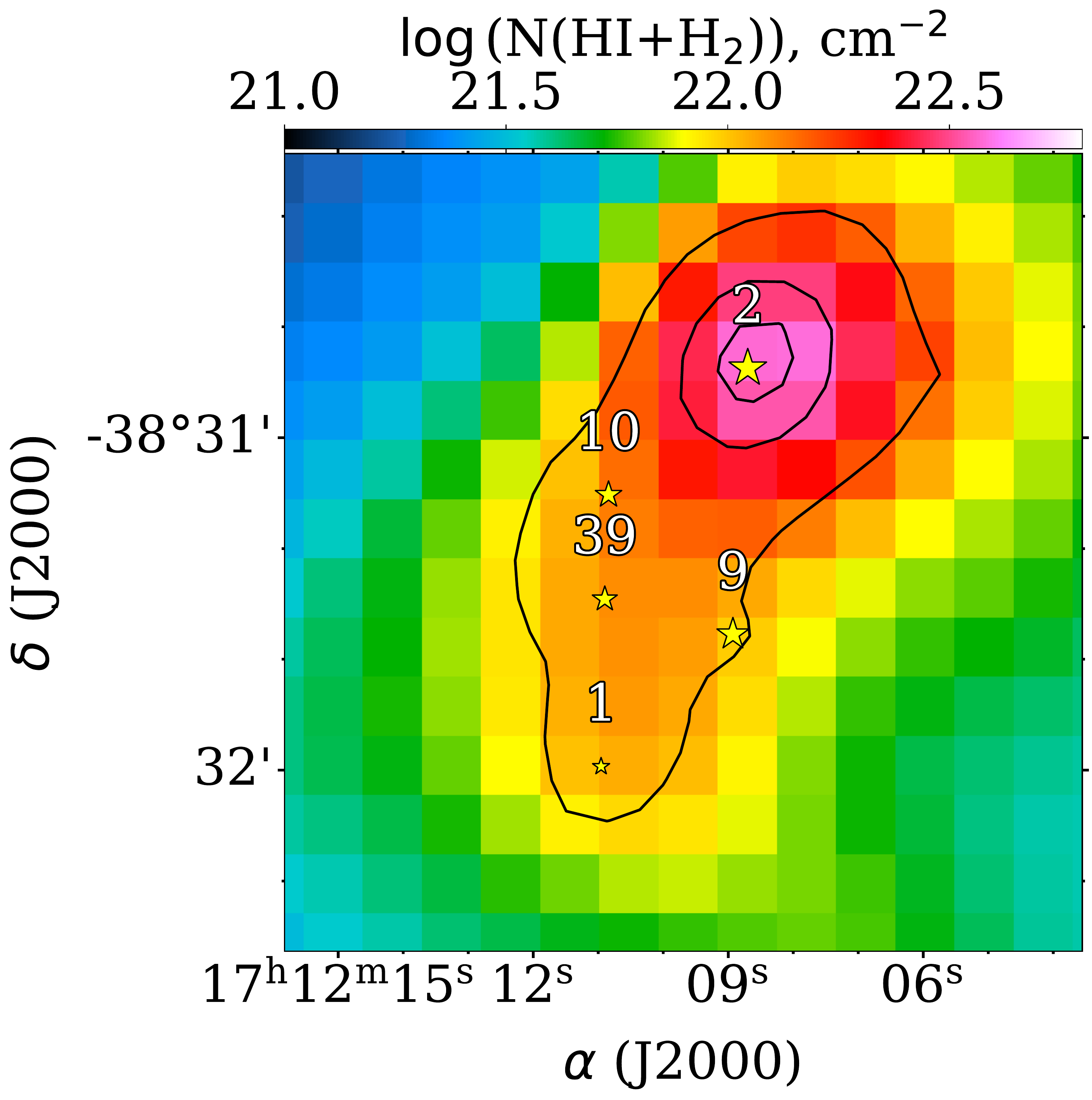}
    \includegraphics[width=0.85\columnwidth,]{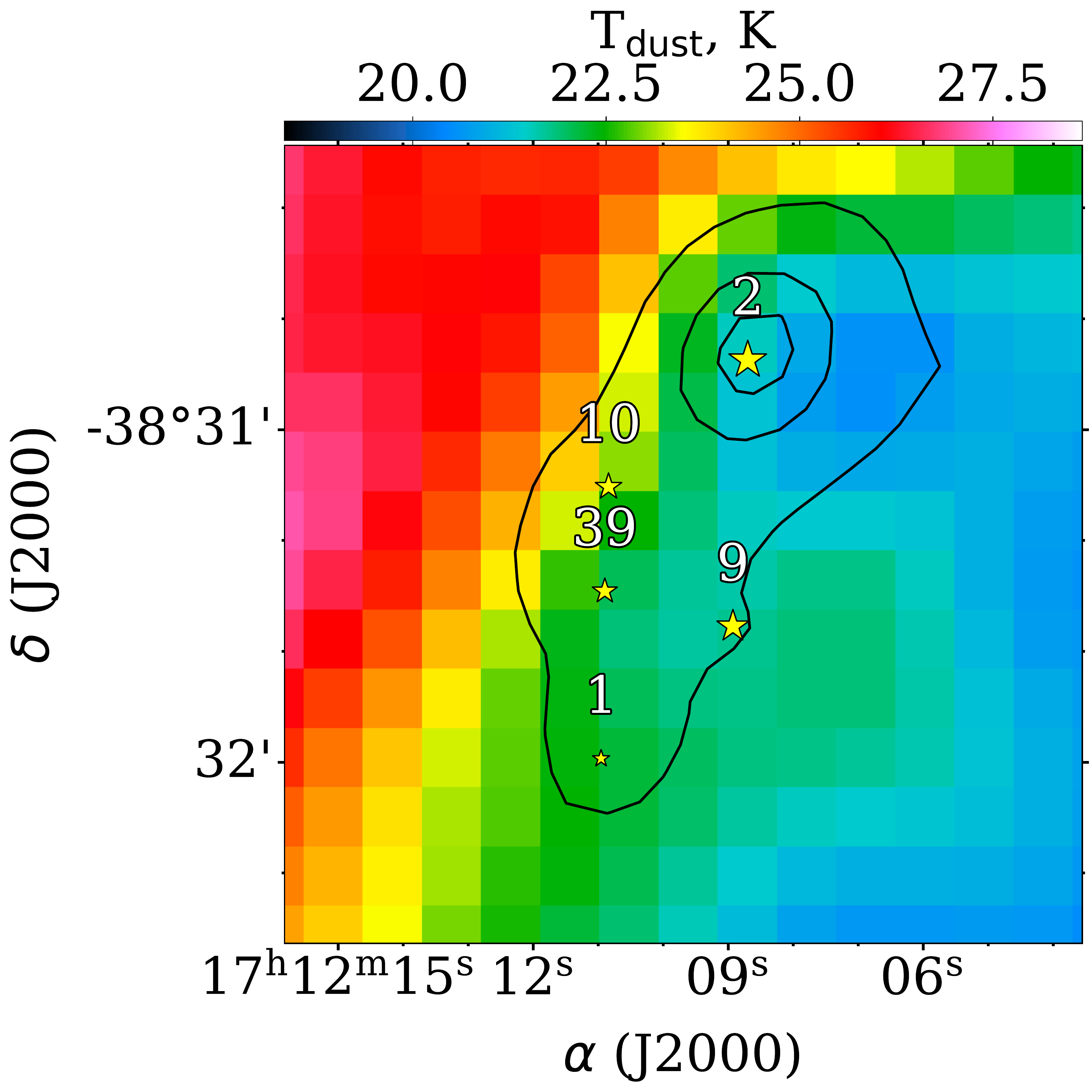}
    	\caption{Column density of hydrogen atoms and dust temperature from the analysis of far-IR dust continuum emission.}
    	\label{fig:farIRDWiebe}
\end{figure}

Fig.~\ref{fig:farIRDWiebe} shows $N{\rm (HI+H_2)}$ value as well as dust temperature in the dense condensation. Obviously, the peak of $N{\rm (HI+H_2)}$ is observed towards the S2~YSO up to $3\times 10^{22}$~cm$^{-2}$. In the directions of the other YSO, $N{\rm (HI+H_2)} \approx 1.0-1.2\times 10^{22}$~cm$^{-2}$. The dust temperature is $\approx 21-24$~K over the whole area of the dense condensation and we consider the $T_{\rm dust}$ as roughly constant there. It only rises up to 27~K towards the PDR and \hii{} region itself.

\subsection{Molecular abundances}

\begin{figure*}    
	\includegraphics[ width=0.5\columnwidth]{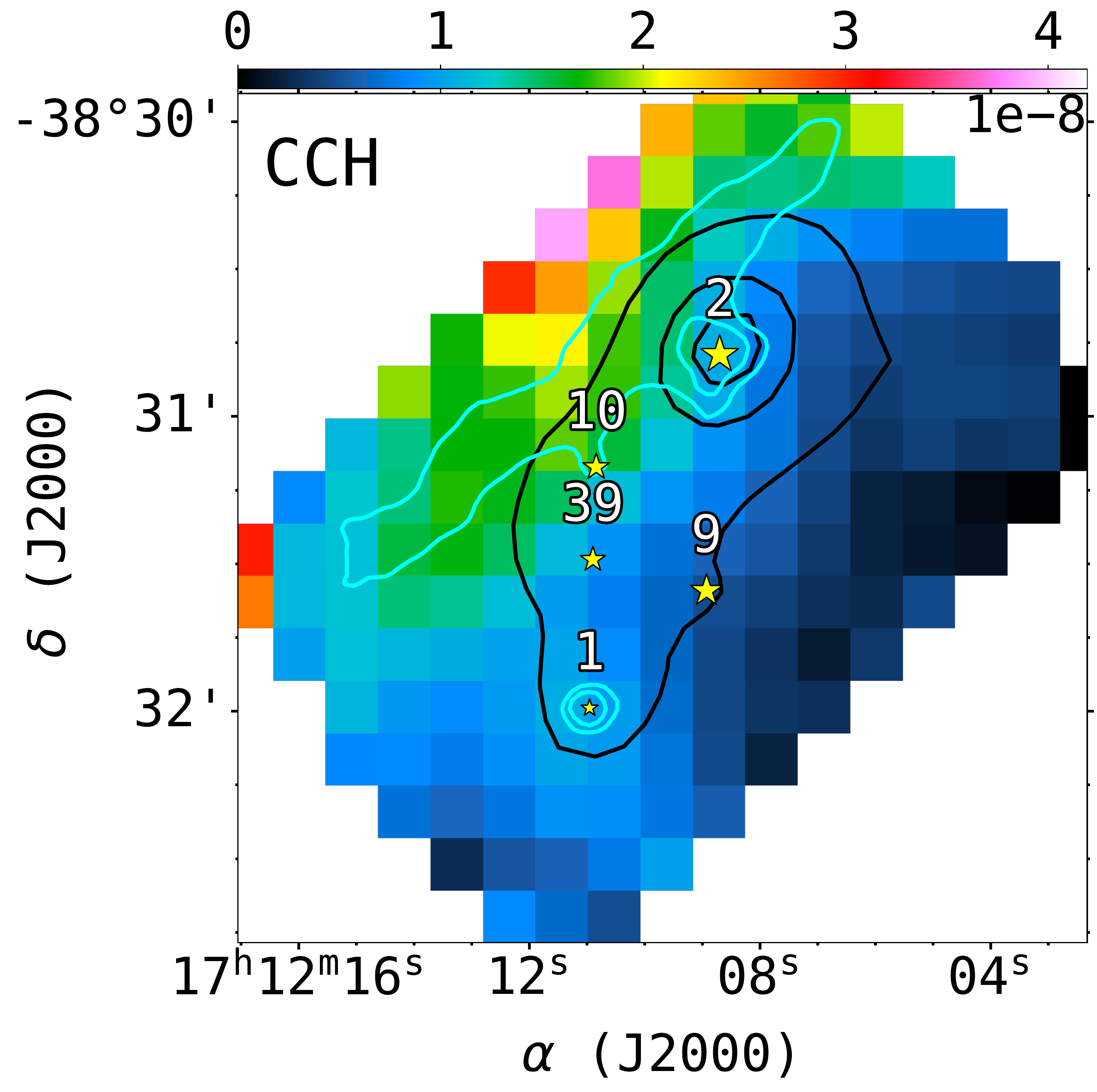}
	\includegraphics[ width=0.5\columnwidth]{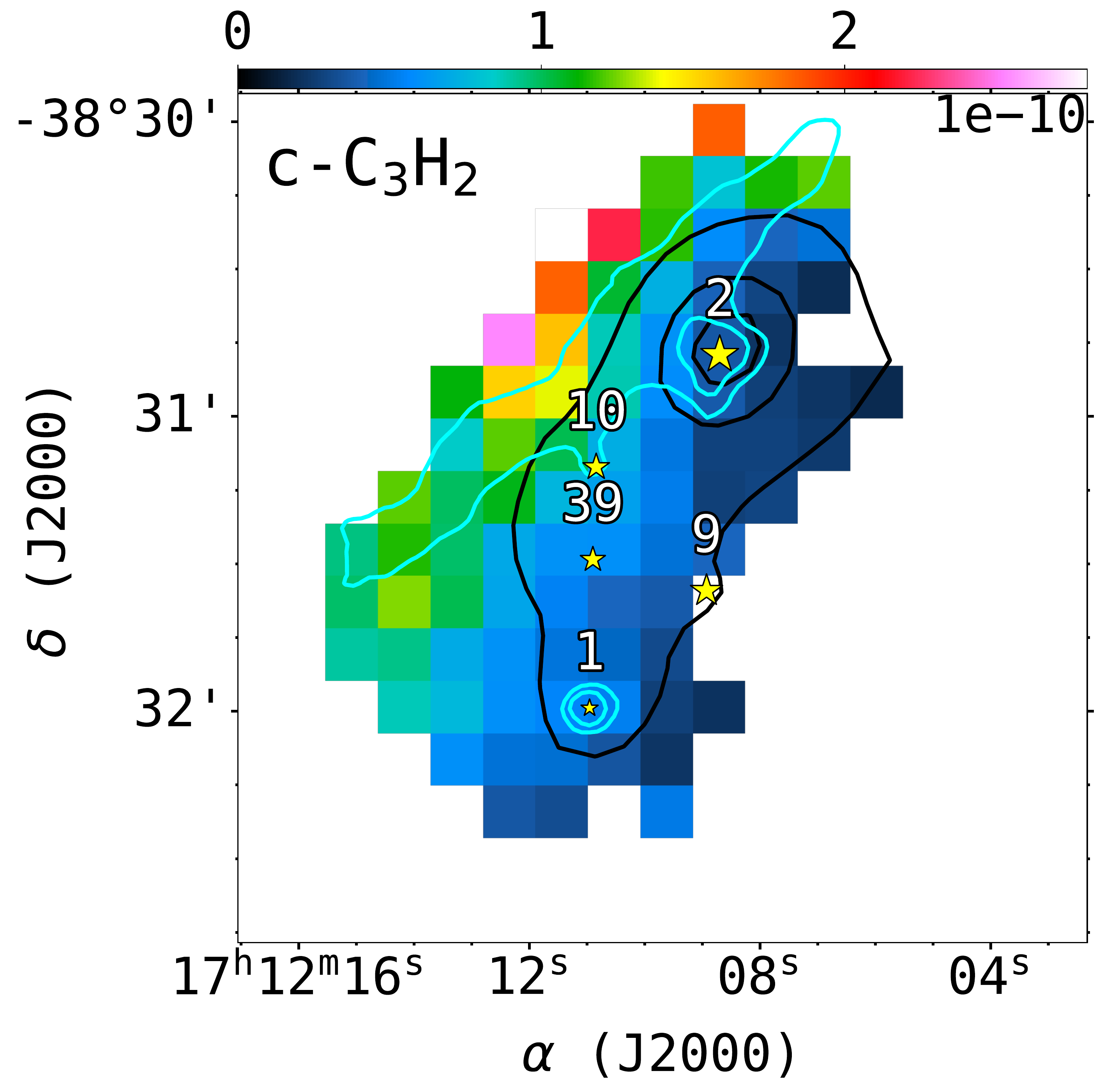}
	\includegraphics[ width=0.5\columnwidth]{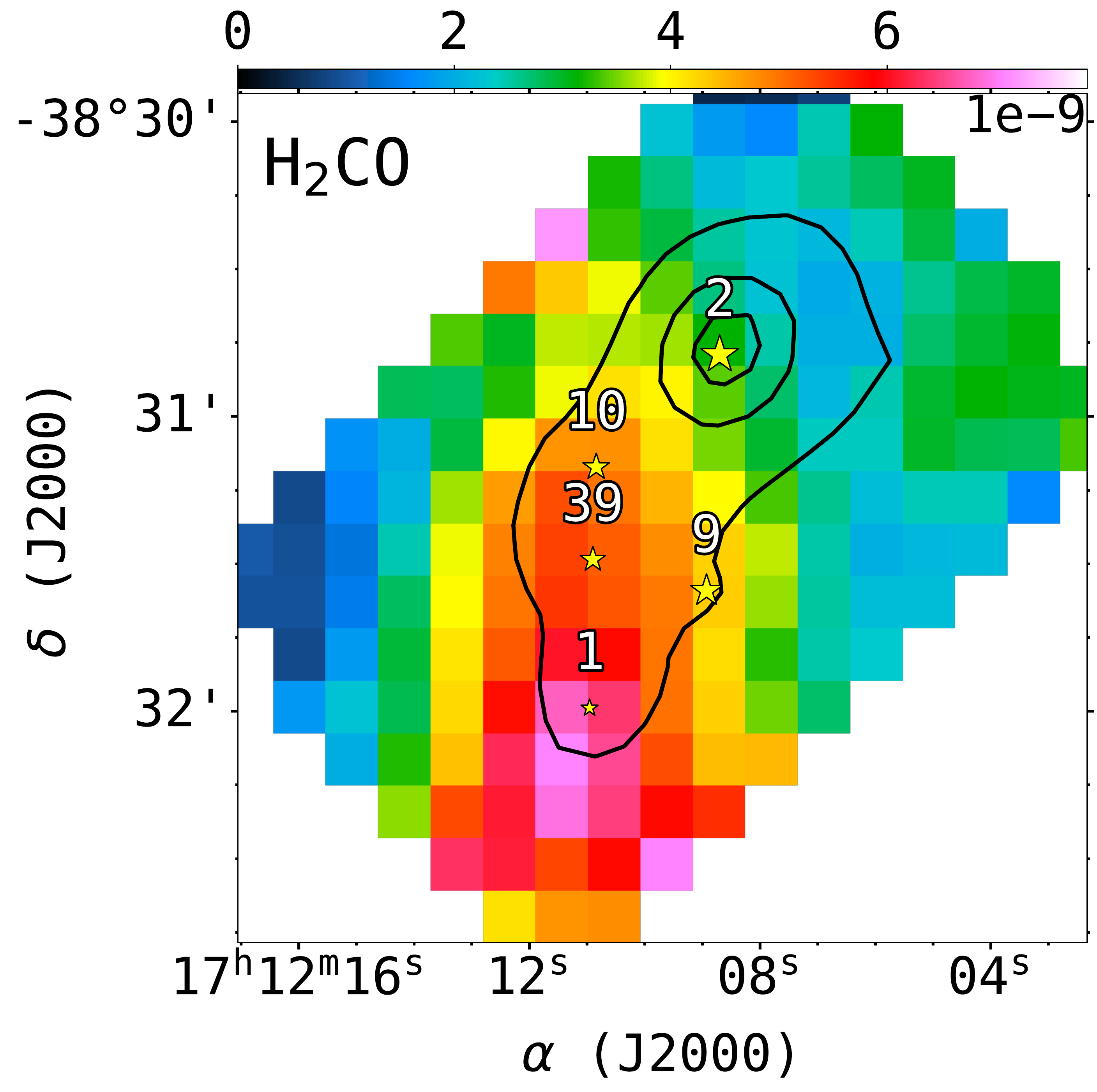}
	\includegraphics[ width=0.5\columnwidth]{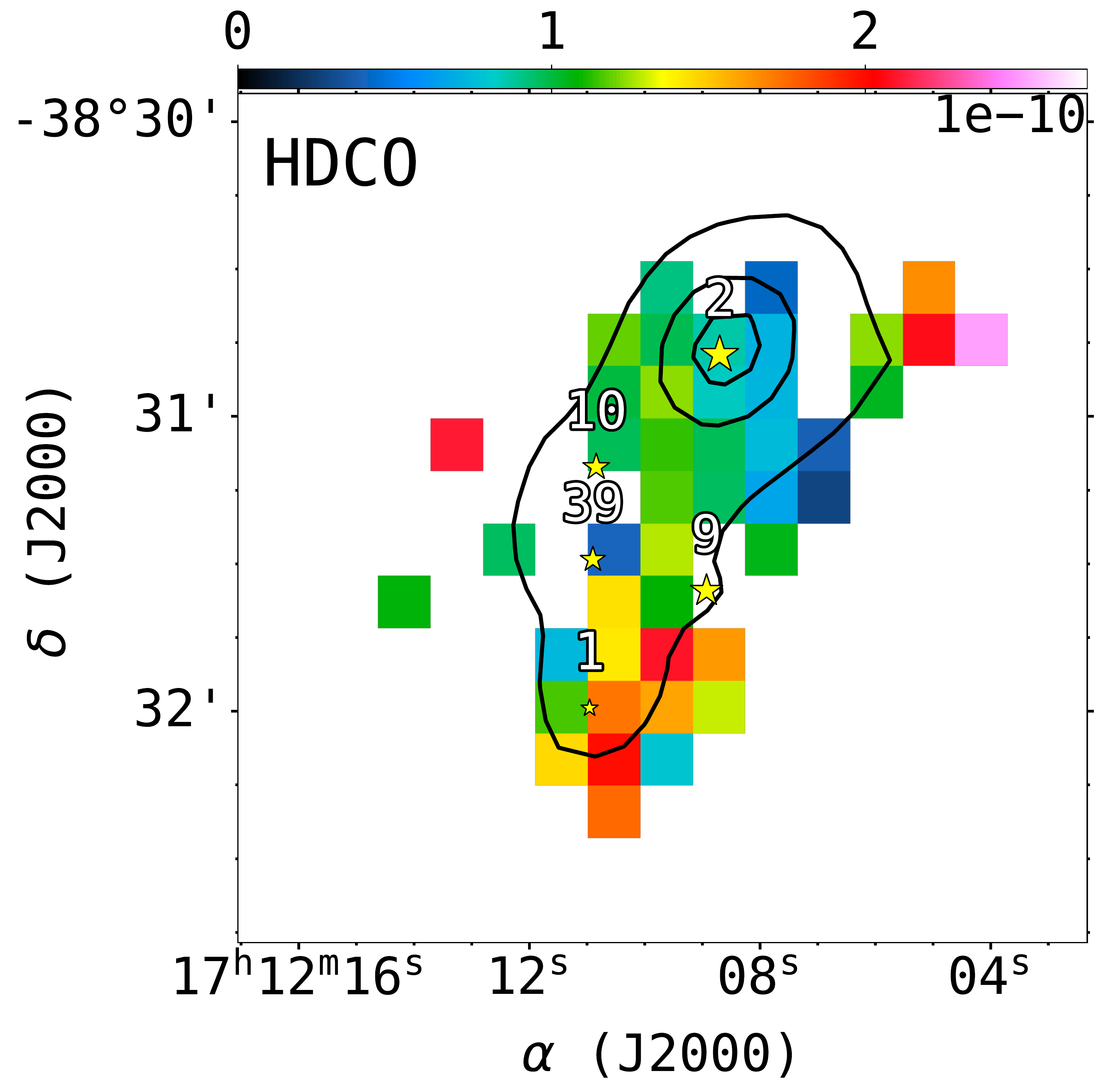}
	\includegraphics[ width=0.5\columnwidth]{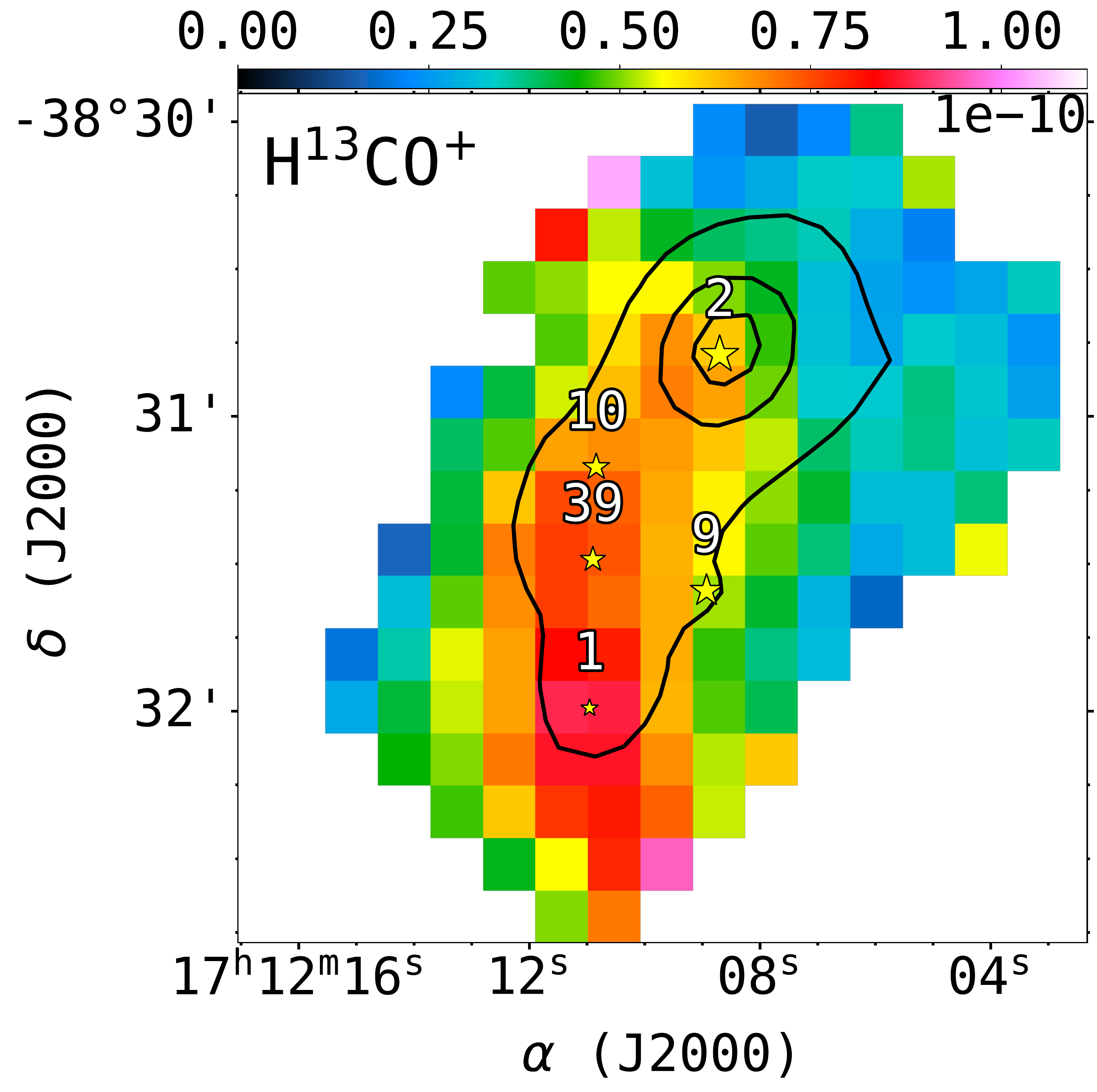}
	\includegraphics[ width=0.5\columnwidth]{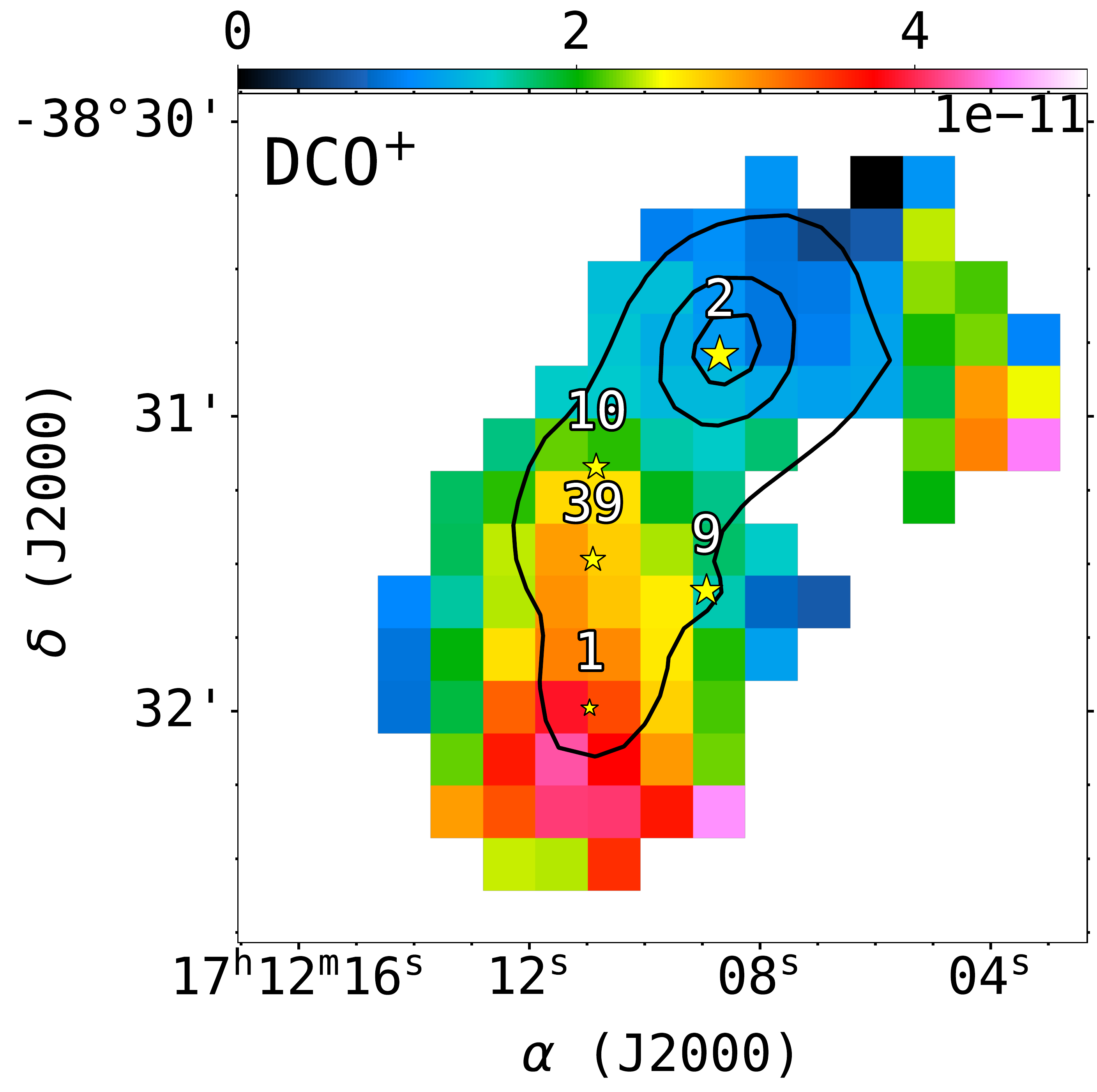}
	\includegraphics[ width=0.5\columnwidth]{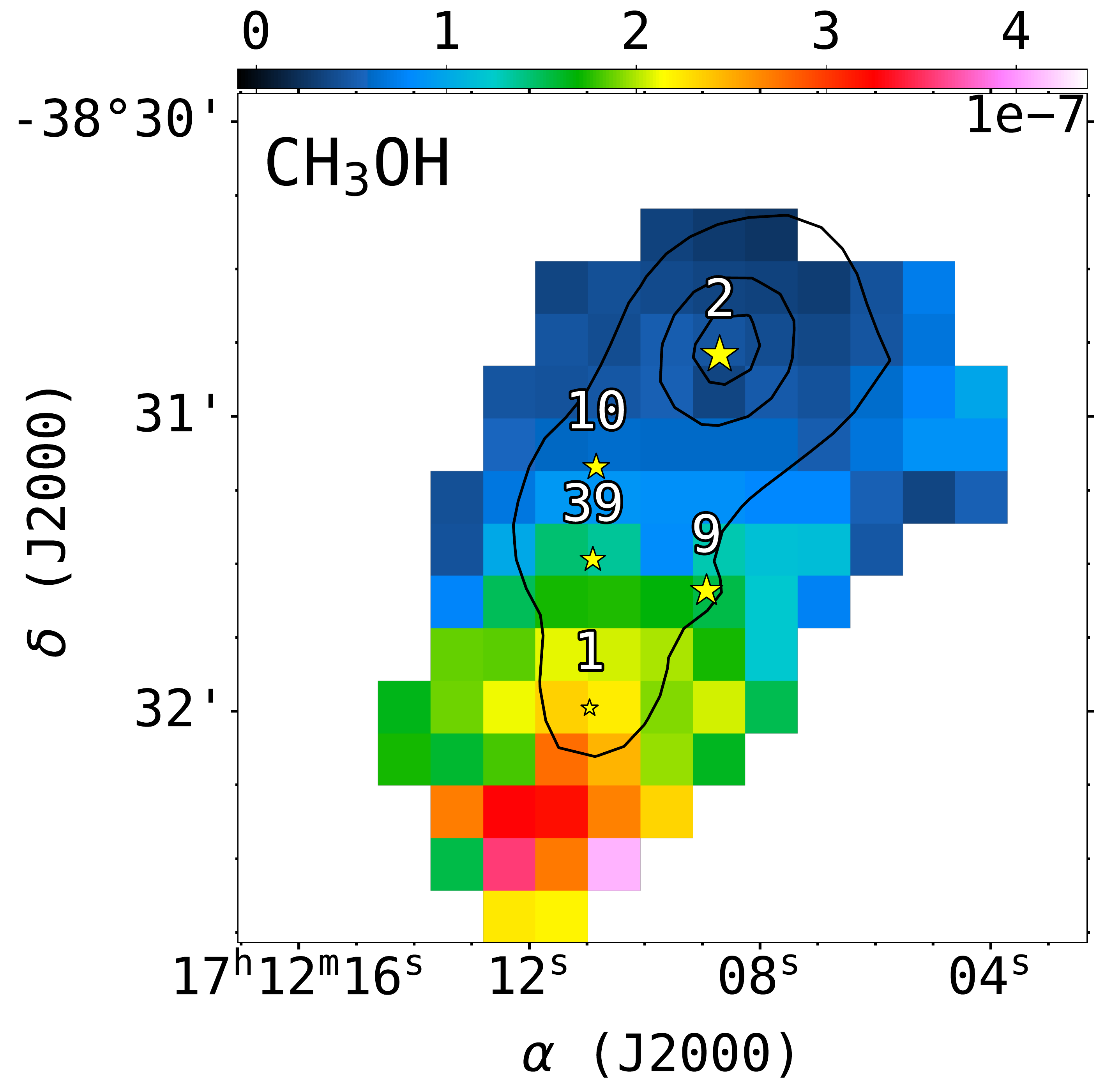}
 	\includegraphics[ width=0.5\columnwidth]{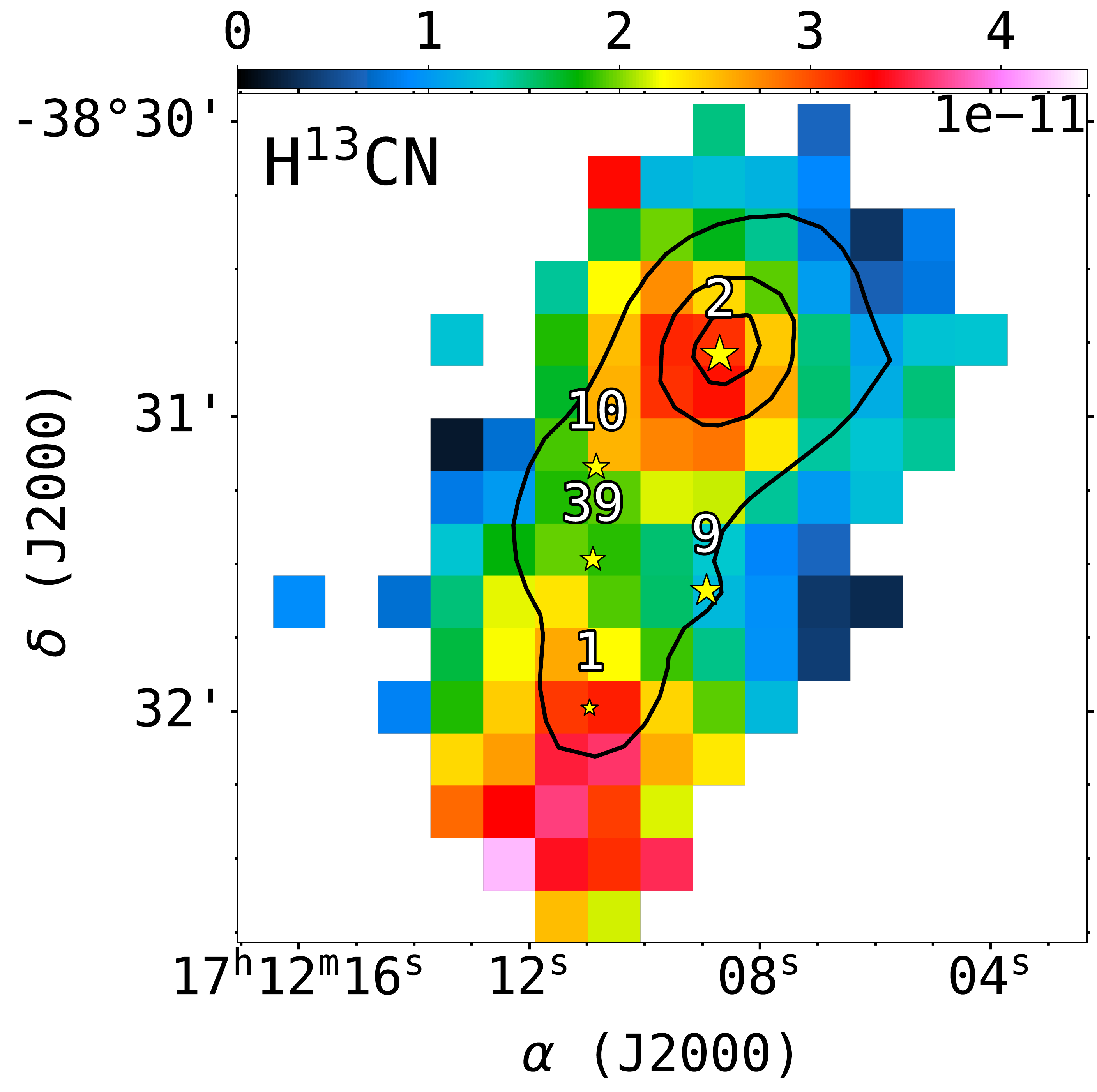}
	\includegraphics[ width=0.5\columnwidth]{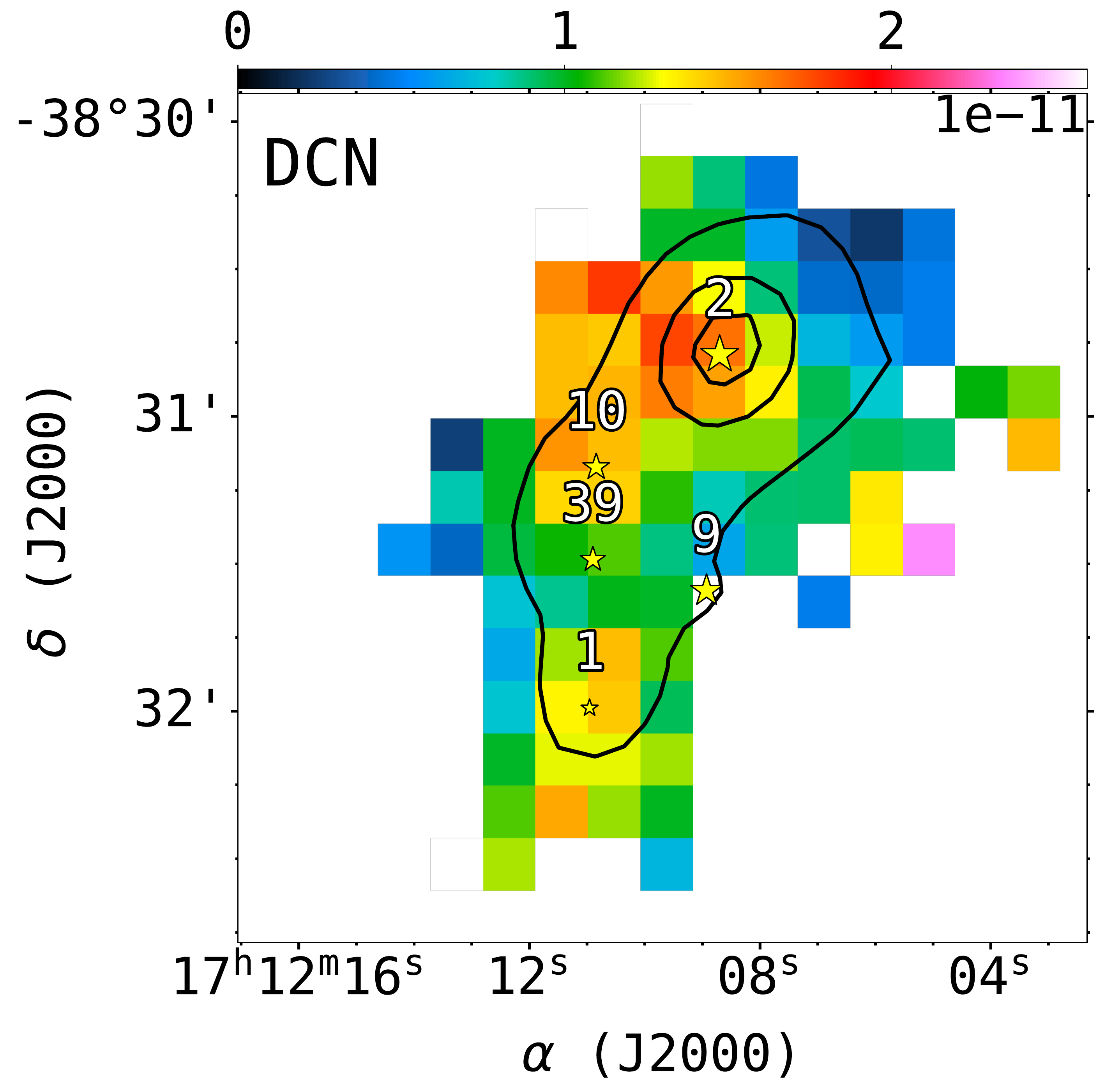}
 	\includegraphics[ width=0.5\columnwidth]{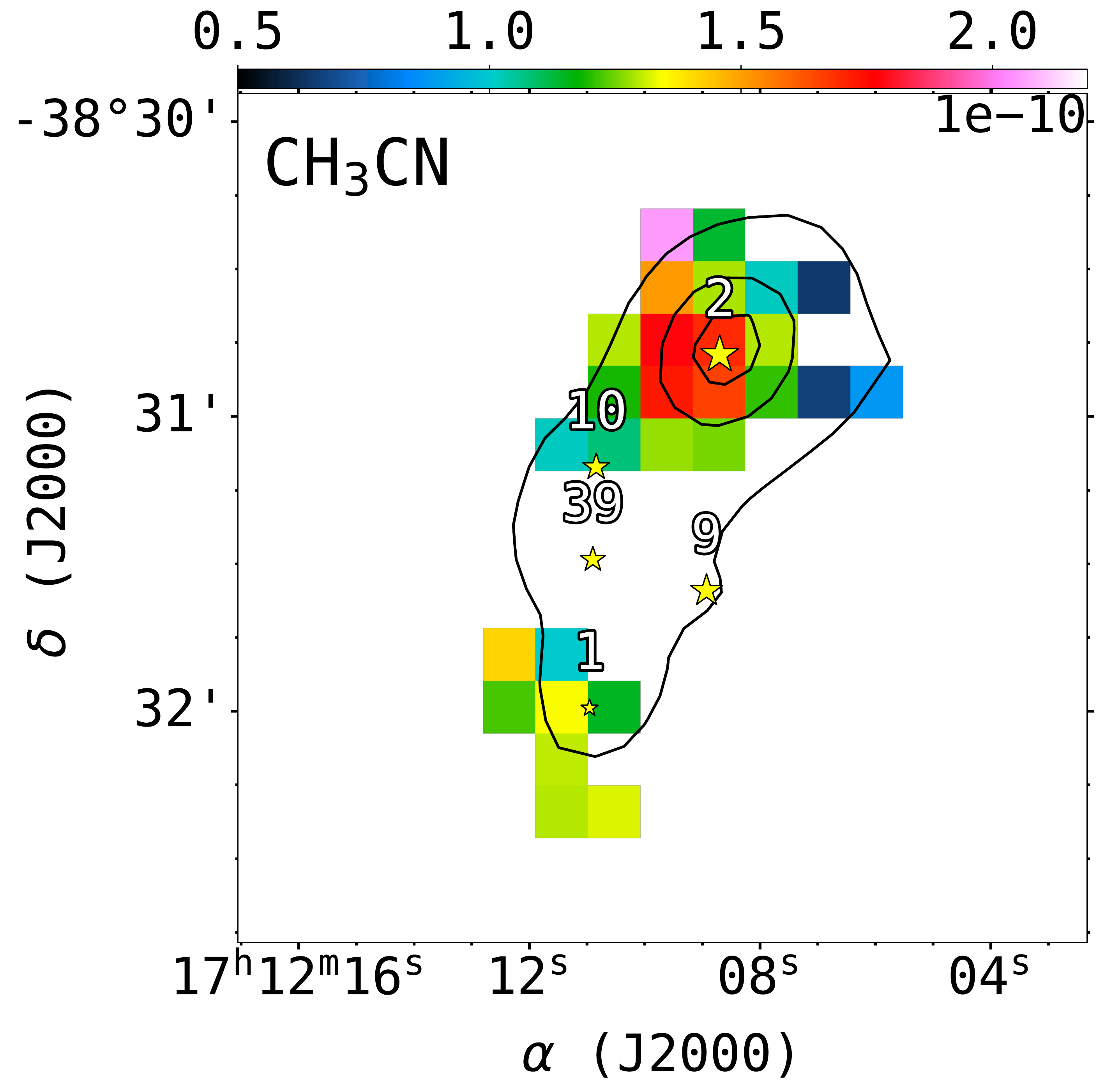}
	\includegraphics[ width=0.5\columnwidth]{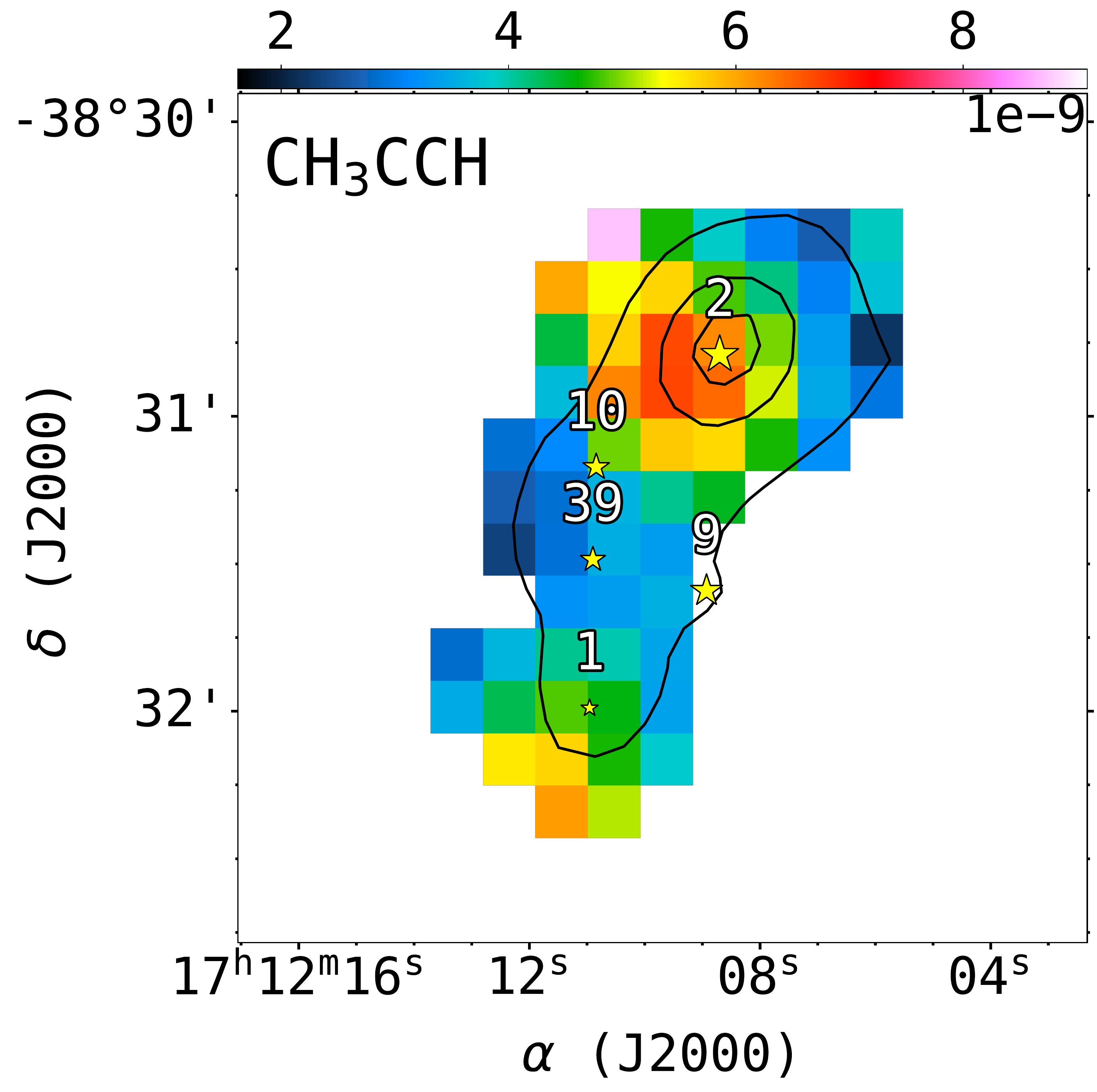}
	\includegraphics[ width=0.5\columnwidth]{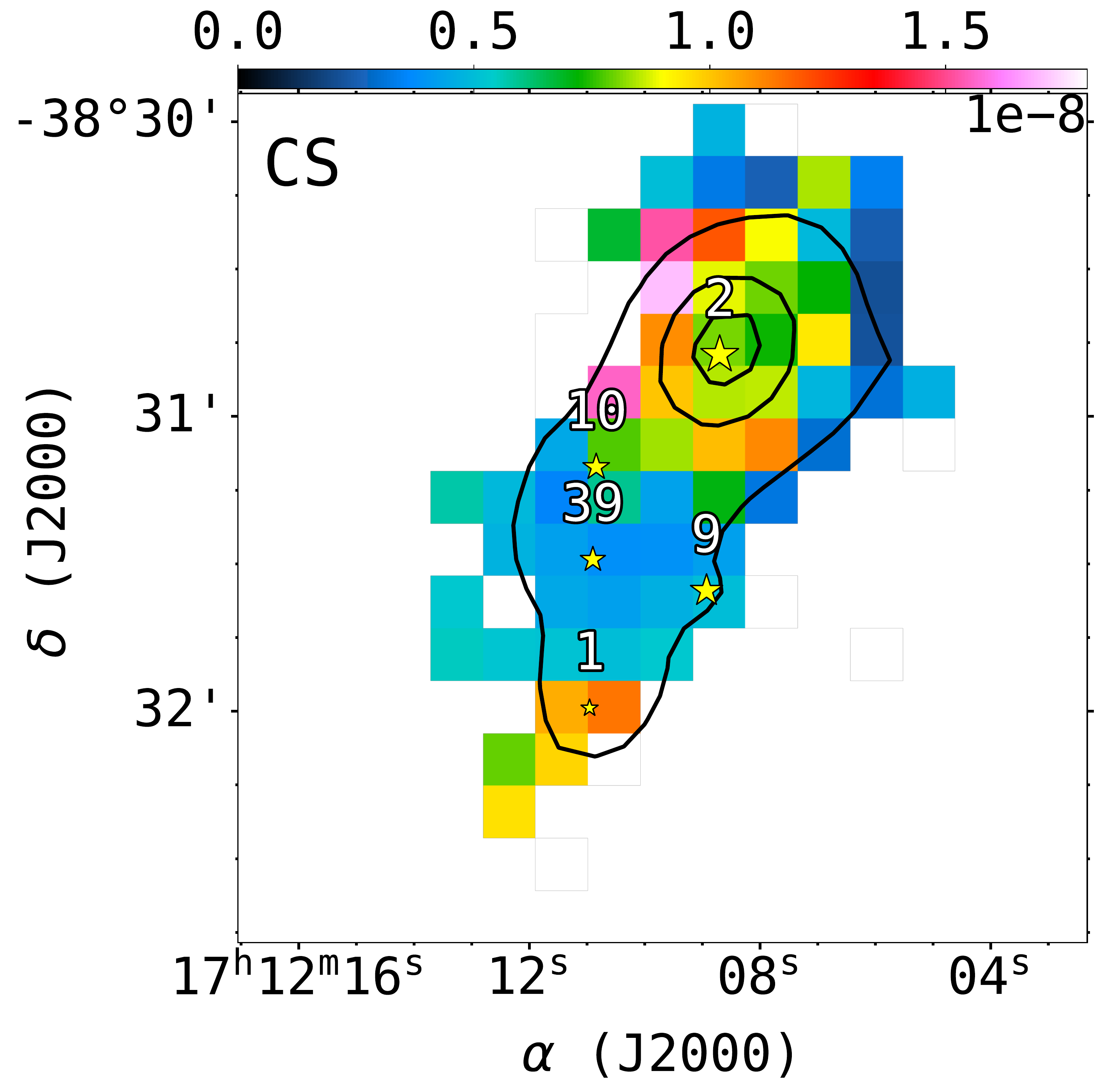}
	\caption{Abundances of molecules. YSOs are marked by yellow stars, the sizes of the symbols are proportional to the YSO masses. The black contours show the emission of dust at 870~$\mu$m, the contour levels are 2.0, 6.0, 10.0 Jy/beam. The blue contours show the emission of dust at 70~$\mu$m, the contour levels are 0.47, 1.0 Jy/pixel.}
	\label{fig:abundances}
\end{figure*}

We show molecular abundances relative to hydrogen nuclei in Fig.~\ref{fig:abundances}. We do not consider CO abundance in the PDR as our LTE analysis can be irrelevant there. CO molecule is abundant up to $2\times 10^{-4}$ in the dense molecular condensation. This value agrees to the carbon elemental abundance observed in diffuse clouds~\citep[see e.~g.][]{2004ApJ...605..272S} but some variations are possible, see discussion in \cite{2020MNRAS.497.2651K}. Therefore, we suggest that almost all carbon is locked in CO molecules in the observed  dense condensation.

Abundances of small hydrocarbons reach their maximum values outside the dense molecular condensation in contrast with all other observed molecules. We see that the region of the highest abundance of small hydrocarbons CCH and c-C$_{3}$H$_{2}$ coincides with the peak of dust continuum emission at 70~\micron{} delineating the RCW\,120~PDR. Abundances of CCH and c-C$_{3}$H$_{2}$ rise there up to $\approx 2.5 \times 10^{-8}$  and $\approx 1.5 \times 10^{-10}$, respectively.

Abundances of such molecules as H$_{2}$CO, HDCO, H$^{13}$CO$^{+}$, DCO$^+$, CH$_{3}$OH reach their maximum values in the direction of S1.  Abundance map for H$_{2}$CO reveals its maximum value, $ \sim 7.5 \times 10^{-9}$, to the south of S1. Similarly, the abundance map for HDCO exhibits a pattern that is reminiscent of the H$_{2}$CO distribution, with the peak abundance $\sim2\times 10^{-10}$ also observed to the south of S1. The abundance map for H$^{13}$CO$^{+}$ shows gradual increase from S2 to the south, reaching its highest value of~$\sim 0.8 \times 10^{-10}$ towards S1. We note that the H$^{13}$CO$^{+}$ abundances multiplied by the $^{12}$C/$^{13}$C ratio can only be considered as a lower limit to the HCO$^{+}$ abundance, because we were not able to take into account optical depth effects. Remarkably, the abundance of H$^{13}$CO$^{+}$ increases farther to the south of S1 up to a factor of 2 comparing with S2. The abundance of DCO$^{+}$ is relatively low, $\approx 1.2\times10^{-11}$, in the vicinity of S2 and then increases almost twice towards S10 and S9 and reaches a maximum value of $\approx 4.3 \times 10^{-11}$ to the south of S1. The CH$_{3}$OH abundance map shares a resemblance with the DCO$^{+}$ abundance map, showing low values near S2, followed by a gradual increase southward. The peak abundance of approximately $4.1\times10^{-7}$ is reached to the south of S1 and is more than 6 times higher than in the vicinity of S2.

Abundances of CH$_{3}$CCH  as well as nitrogen-bearing molecules H$^{13}$CN, DCN and CH$_{3}$CN show two peaks in contrast with all the other ones. Both peaks are related with the S1 and S2. For HC$^{13}$N we observe the maximum abundance of $1.5\times10^{-11}$ to the south of S1. The second maximum towards S2 is lower by a factor of 1.5. Again we note that the H$^{13}$CN can only give a lower limit for the main isotopologue HCN due to the unaccounted optical depth effect. Examining the DCN abundance map, we see that this molecule exhibits its highest abundance around S2, with a value of $\approx 1.8 \times 10^{-11}$. Additionally, DCN shows a relatively constant abundance level to the east of S2. Another notable peak in DCN abundance is observed in the vicinity of S1 at the same level. The CH$_{3}$CN abundance map exhibits its highest values of $1.8\times10^{-10}$ towards S2. Towards S1, the abundance is slightly lower, $1.4\times10^{-10}$. As we move to the south from S1, the abundance does not rise much, maintaining almost the same values. The abundance map for CH$_{3}$CCH reveals its peak towards S2, where it reaches a maximum of $6.5\times10^{-9}$. The abundance is roughly 1.5 times lower towards S1. In contrast to  H$^{13}$CN, DCN and CH$_{3}$CN,  as we move farther south, the abundance of  CH$_{3}$CCH increases, eventually reaching values similar to those observed towards S2.

CS abundance is $\approx 0.5 \times 10^{-8}$ to the west of S2, experiences a modest increase to $1.1 \times 10^{-8}$ on the southern side, and reaches $ \approx 1.5 \times 10^{-8}$ to the north-east of S2. In the vicinity of S1 we observe CS abundances around $1.0 \times 10^{-8}$, which remain almost constant to the south-west.  It is important to note that the emission between S1 and S2 appears to be optically thin, preventing us from determining the excitation temperature in this region, therefore we applied the average excitation temperature of 9~K.

Concluding with abundances, we note that oxygen-bearing molecules H$_{2}$CO, HDCO, H$^{13}$CO$^{+}$, DCO$^{+}$ and CH$_3$OH have their abundance maxima towards and to the south of the S1 YSO. Molecules without oxygen, DCN CH$_3$CN, CH$_3$CCH and CS, have their main maxima towards the S2 YSO and secondary maxima towards the S1 YSO. Small hydrocarbons CCH and c-C$_{3}$H$_{2}$ are mostly abundant towards the border of the PDR. We plan to discuss this result in our next study. It is possible that the observed chemical structure is related not only to the dust temperature and thermal desorption, but another processes like shock waves and non-thermal desorption.

\subsection{Deuterium fractination}

Among all the D/H ratios derived in this study from different deuterium-bearing species, HDCO/H$_{2}$CO stand out with the highest deuterium fractionation, which remains mostly uniform with a value of $0.02-0.04$. The DCO$^{+}$/HCO$^{+}$ ratio gives the lowest value $\leq 0.01$. The minimum DCO$^{+}$/HCO$^{+}$ ratio of 0.005 is found at S2 and the maximum of 0.01 is observed to the south of S1. Deuterium fraction of DCN/HCN is 0.005 throughout the considered molecular condensation.

\subsection{Gas kinematics}
\label{sec:PV_diag}
\begin{figure*}
    \includegraphics[width=2\columnwidth,]{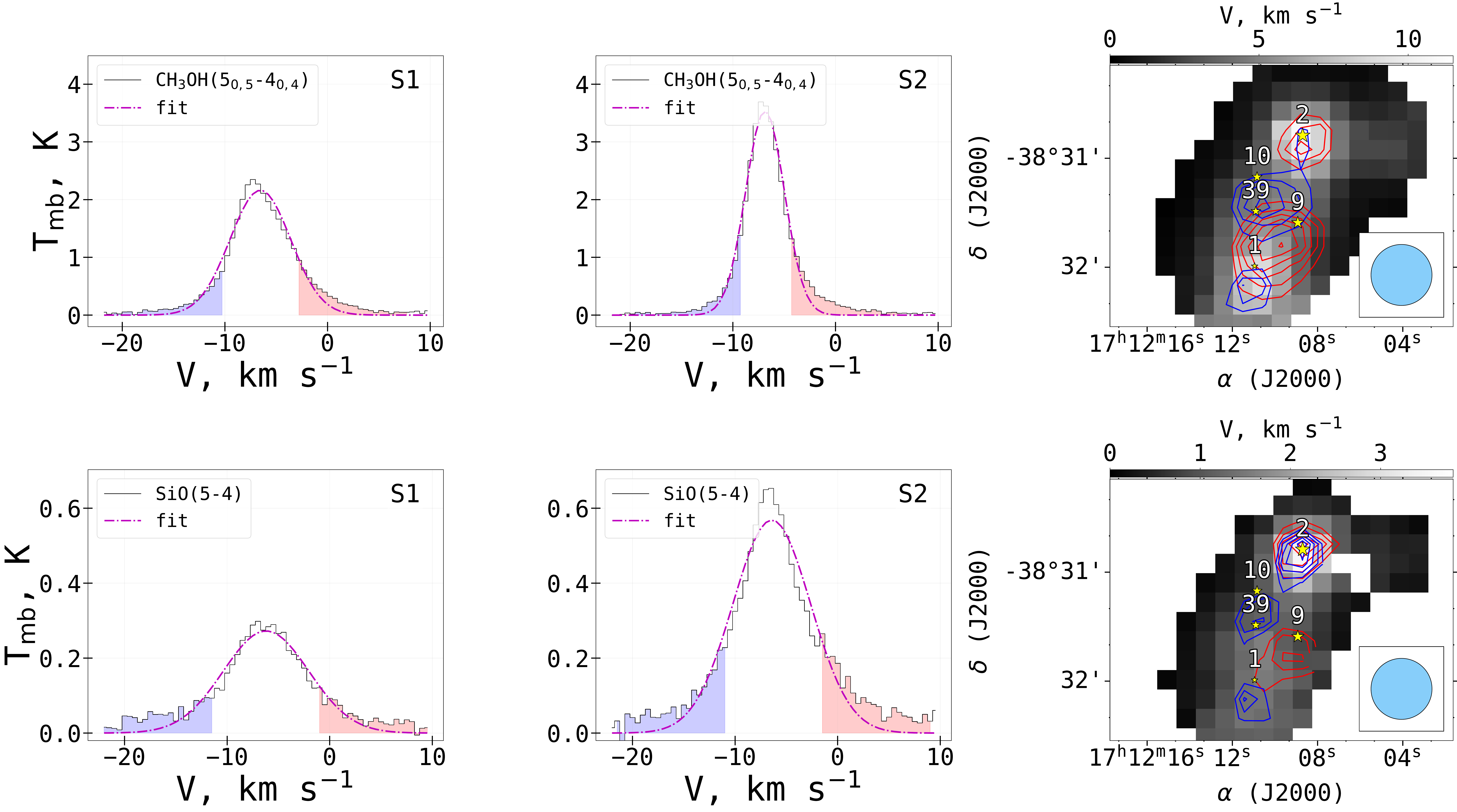}
    	\caption{Upper panel: The background grey-scale represents the integrated intensity of the middle component of CH$_3$OH~(5$_{0, 5}$--4$_{0, 4}$, A$^{+}$) line emission. The blue and red contours indicate the blue- and red-shifted CH$_3$OH line emission, respectively. Contour levels for the redshifted gas are 20, 24, 28, 32, 36, 40~$\sigma$, while contour levels for the blue-shifted gas are 14, 16, 18, 20~$\sigma$ ($\sigma$ = 0.15 K*km/s)
    	Lower panel: The background grey-scale represents the integrated intensity of the middle component of SiO~(5-4) line emission. The blue and red contours indicate the blue- and red-shifted SiO line emissions, respectively. Contour levels for the redshifted gas are 6, 7, 8, 9, 10~$\sigma$, while contour levels for the blue-shifted gas are 5, 5.5, 6, 6.5, 7, 7.5~$\sigma$ ($\sigma$ = 0.14 K*km/s).}
    	\label{fig:wings}
\end{figure*}

In order to study outflows around the YSOs, we plot maps of the SiO~(5--4) and CH$_3$OH~(5$_{0, 5}$--4$_{0, 4}$) line emission in the red, middle and blue intervals, see Sec.~\ref{sec:methodkinem}. Examples of the intervals for S1 and S2 YSOs and the maps of the emission in the intervals are shown in  Fig.~\ref{fig:wings}. We find the red and blue wing emission in both lines almost intersecting in S2. Therefore, we propose that the outflow in this YSO is oriented along the line of sight. At the same time, we see that the blue wing is less prominent in the methanol emission compared with SiO as it can also be seen in the example of the SiO and methanol spectra on the left of the plot. There are red and blue wings around S1. This emission looks as a bipolar outflow oriented from north-west to south-east. There is another area of the blue wing emission towards S39. Methanol profiles look as consisting of several overlapping components, but separation of these components in some particular area is impossible. The possible reason is our beam size, which is comparable to the angular separation between the YSOs.

\begin{figure} 
\hspace*{-5mm}
\includegraphics[width=1.1\columnwidth]{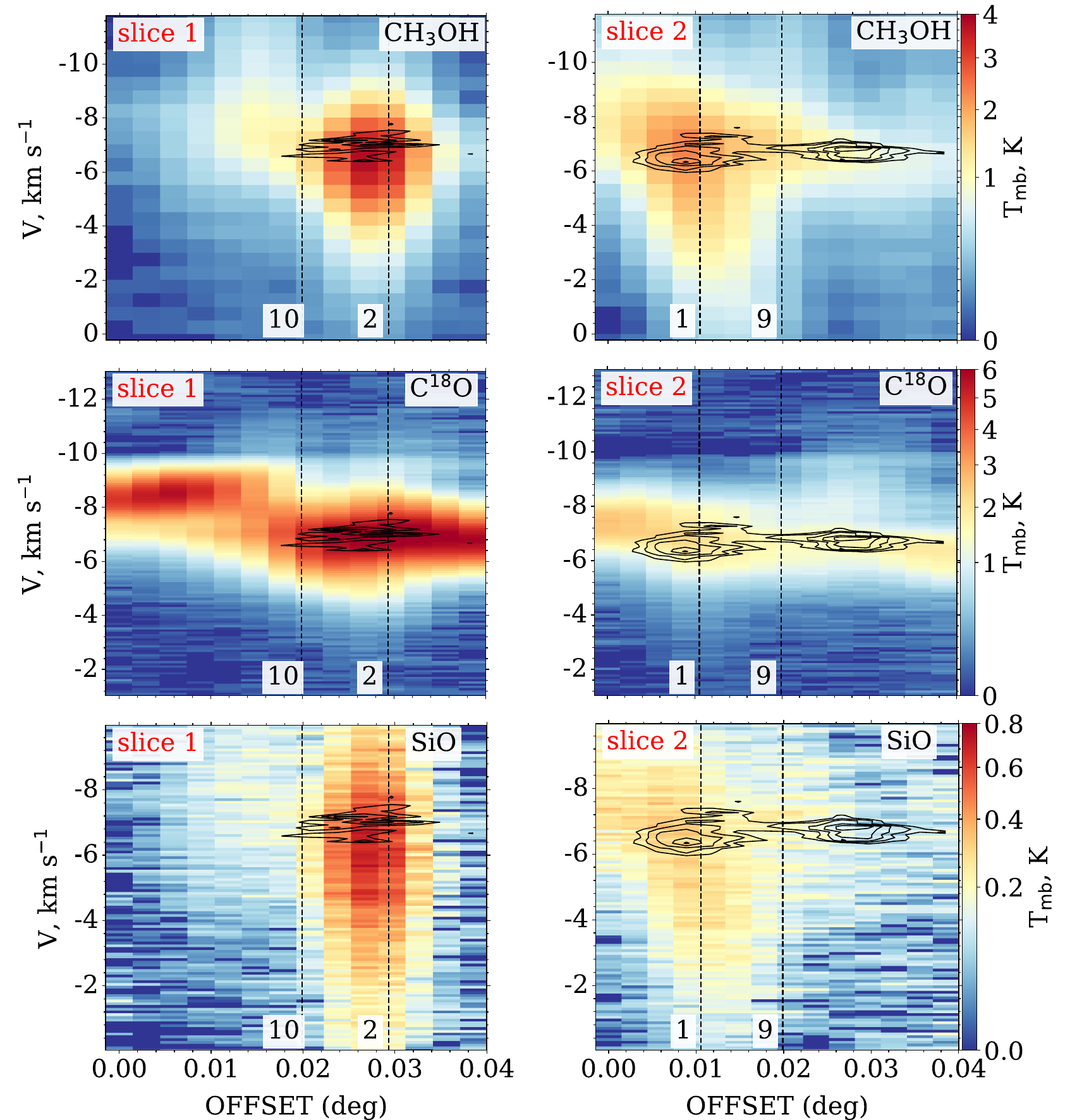}
\caption{PV~diagrams of  CH$_3$OH~(5$_{0, 5}$--4$_{0, 4}$), SiO~(5--4) and C$^{18}$O~(3--2) line emission shown by colour. The black dashed lines indicate the positions of YSOs. PV~diagram of DCO$^{+}$~(3-2) line is shown by black contours. The contour levels are from 0.25~K to 0.6~K with a step of 0.05~K. }
\label{fig:pos-vel-diag}  
\end{figure}

In order to explore kinematics on a larger scale,  we plot position-velocity (PV) diagrams from cuts, which oriented parallel to the RCW\,120 PDR using the CH$_{3}$OH~(5$_{0, 5}$--4$_{0, 4}$), C$^{18}$O(2--1) and DCO$^{+}$~(3--2) lines in Fig.~\ref{fig:pos-vel-diag}. The cuts itself shown as white arrows in Fig.~\ref{fig:integrated_intensity_map} overlaid on ${^{13}}$CO panel and labelled as ‘slice~1’ and ‘slice~2’. These cuts intersect S10 and S2 (referred to as slice~1), as well as S1 and S9 (referred to as slice~2). A noticeable contrast in velocity components emerged when examining the position-velocity (PV) plot along slice~1 of C$^{18}$O (Fig.~\ref{fig:pos-vel-diag}). Moving along the slice from south-east to north-west, we clearly see how the line peak velocity shifts from approximately --9\kms{} to --6\kms{} close to S10. There is no dense gas along this slice to the left from S10 as we can see in Fig.~\ref{fig:farIRDWiebe}. This shift becomes less pronounced at the slice\,2, which is away from the PDR. We find almost straight line along the slice\,2 with a minor feature of the self-absorption.

PV~diagrams of the lines of such species as CH$_3$OH and SiO are different from what we observed for C$^{18}$O. However, they share a notable similarity: a broad line, associated with the outflow, is observed towards S2. Comparing directions of both S1 and S2 YSOs, we find that the outflow is more prominent in the SiO diagram than in the CH$_3$OH diagram. 

However, the PV~diagram of the DCO$^+$(3--2) line is narrow even towards the YSOs. We see a narrow line between S2 and 10 at the slice\,1 and no DCO$^+$ in the region with the shifted C$^{18}$O(2--1) line. There are two separated regions with the same velocity at the slice\,2, namely, the first between S1 and S9 and the second with no YSOs.  Positions of the YSOs coincide with the borders of the regions on both diagrams. Obviously, the outflows are not visible at the DCO$^+$(3--2) diagrams, but only a quiescent molecular gas in contrast with all molecules considered above.
    
\section{Discussion}
Analysis of the gas kinematics and abundances of molecules shows that the dense molecular condensation on the border of the RCW\,120 PDR consists of two parts: the southern part containing S1 and the northern one containing the S2 YSOs. Both contain quiescent molecular gas visible through the narrow DCO$^+$(3--2) line and outflows visible through the broad CH$_3$OH and SiO lines. The narrow C$^{18}$O line on the PV~diagram agrees with the velocity of a layer shocked by the expanding \hii{} region when we compare our results with PV~diagrams made by \citet{2023MNRAS.520..751K}. \citet{Luisi_2021} found that the PDR expands towards the observer. We see the blue C$^{18}$(2-1) velocity shift from --6\kms{} to --9\kms{} near the S10 YSO. The note that the shift can be related with different density of the molecular condensation where the dense gas around S2 decelerates the shock more effectively than the less dense gas to the east from the YSO.

The southern and the northern parts of the condensation have different molecular content, where nitrogen-containing molecules (H$^{13}$CN, DCN, CH$_3$CN) are more abundant in the northern part but oxygen-containing molecules (DCO$^{+}$, SiO, H$_{2}$CO, CH$_3$OH) are more concentrated in the southern part. Small hydrocarbons (CCH and c-C$_{3}$H$_{2}$) are concentrated in the direction of the PDR. This phenomenon can be attributed to the gas-phase formation of c-C$_{3}$H$_{2}$, which is considered to originate from CH$_{4}$, through the ``Warm Carbon Chain Chemistry'' (WCCC) \citep[see e.~g.][]{Hassel_2008, 2013ChRv..113.8981S}. The presence of UV radiation causes the dissociation of CO molecules, resulting in the formation of carbon atoms, thereby contributing to these enhanced abundances.

In spite of almost the same dust temperature $T_{\rm dust} \approx 21-23$~K in all YSOs, there are distinct chemical differentiation in the dense molecular condensation. Obviously, these differences are not caused by evaporation of dust grain mantles. CO molecules observed in the gas phase at the level of $\sim 10^{-4}$ are evaporated from the dust grains since their evaporation temperature is less than 20~K at the densities in our YSOs \citep[see ][]{2015A&A...582A..41H}. 

Excitation of such COMs as CH$_3$CCH and CH$_3$CN line emission shows higher values of $T_{\rm rot}$ than $T_{\rm dust}$. Analysis of the methanol emission also shows that the excitation of lines can take place at higher temperatures as the $T_{\rm k} \geq T_{\rm dust}$. Therefore, excitation of these COMs occurs not in a cold medium traced by the far-IR emission from {\it Herschel}  \citep[see observations by][]{2016AA...591A.149M}, but in warmer and more compact regions around the YSOs. \cite{Figueira_2019} resolved S2 into five fragments using the {\it ALMA} interferometer. Therefore, compact components exist there. Our values for $T_{\rm rot}$ for CH$_3$CN in the S2 appears to be consistent with the values previously found by \cite{Kirsanova_2021} while they used the $J_{K} = 12_{K}-11_{K}$ line series only. They also suggest that hot gas exist at least around S2. We do not analyse the CH$_3$CCH and CH$_3$CN line emission with $K>3$ here leaving this issue for forthcoming papers but concentrate ourselves on large-scale distribution of molecules. Our analysis of the methanol emission lines is also does not contradict to \cite{Kirsanova_2021}. Both these studies reveals that filling factor $f<1$ towards the YSOs. Comparing excitation temperatures for CH$_3$OH, CH$_3$CCH and CH$_3$CN, we propose that methanol appears in colder and in more extended area around YSOs than CH$_3$CCH and CH$_3$CN. Moreover, excitation of CH$_3$CN takes place in more warm and dense regions comparing with CH$_3$CCH. CH$_3$CN and CH$_3$CCH are also more abundant in the high density region around S2. Therefore, we suggest an onion-like chemical and physical structure around the S1 and S2 YSOs. Sequential appearance of CH3OH to CH3CCH and CH3CN in the gas around YSOs means transition to warmer gas closer to the YSO.

\section{Conclusions} 

Using the APEX telescope, we performed observations of molecular line emission in the dense condensation on the border of the RCW\,120~PDR, where YSOs are still embedded into parental molecular gas. We covered frequencies from 200 to 260~GHz. Overall, we identified 20 molecules which produce the brightest line emission, including diatomic molecules like CS and SO, as well as complex organic molecules like CH$_3$CN and CH$_3$CCH. We made the following conclusions.

\begin{itemize}
    	\item We found that all the integrated intensity maps can be divided into three groups: (i) Molecules, that are ubiquitous to PDR ($^{13}$CO, C$^{18}$O, small hydrocarbons CCH and c-C$_3$H$_2$), (ii) {\bf ones} showing spatial correlation of their line emission with the dust continuum at 870~$\mu$m (CS, CH$_3$OH, SiO, DCO$^+$, H$^{13}$CO$^+$,  H$_2$CO, HNCO, HDCO), (iii) {\bf ones}, whose emission predominantly appears toward S2 YSO (sulphur-bearing species: SO, C$^{34}$S, H$_2$S, H$_2$CS; nitrogen-bearing species: DCN, H$^{13}$CN, CH$_3$CN; and CH$_3$CCH).
   	 
    	\item Physical conditions were studied under the LTE assumption using a number of transitions of CH$_3$CN, CH$_3$CCH, CH$_3$OH and SiO. Analysing the CH$_3$CCH data, we found that the rotational temperature for S1, S2, and S10 YSOs is around 40~K. Analysis of CH$_3$CN shows higher values of rotational temperatures towards S1, S2, and S10 up to 61~K. Non-LTE analysis of methanol emission showed that gas temperature was from 20 to 40~K in the YSOs. S2~YSO appears as the densest region. S1~YSO is the coldest YSO with the highest methanol column density.
   	 
    	\item Analysis of spectral lines from shock tracers such as CH$_3$OH and SiO revealed broad, asymmetric profiles in the vicinity of the YSOs. This implies a presence of outflows within the observed regions. We propose that there is an outflow along the line of sight towards S2, and we also infer the presence of a bipolar outflow in the vicinity of S1. Further studies of CH$_3$OH and SiO emission are needed to ascertain origins of the outflows because our data have insufficient spatial resolution.
   	 
    	\item Maps of molecular abundances could be also divided into three groups: (i) The highest abundances is observed near the PDR (CO, CCH, c-C$_3$H$_2$), (ii) The highest abundances are observed in the vicinity of S1, (iii) The abundances are about the same in the vicinity of S2 and S1 or higher in S2. Abundances of small hydrocarbons, specifically CCH and c-C$_{3}$H$_{2}$, are related to group (i), revealing a concentration of these molecules towards the PDR. 
   	 
    	\item The deuterium fractionation, with a peak value of approximately 3\%, was found for the HDCO/H$_{2}$CO pair of molecules in S2~YSO, highlighting the influence of the well-known kinetic isotope effect. Conversely, the lowest deuterium fraction, falling below 0.5\%, was found for the DCO$^+$/HCO$^+$ pair in S2. These results will be useful for further astrochemical modelling.
     
         \item The enhanced production of hydrocarbons in the northern part of the dense condensation is the only one influence of the feedback from the massive star on the molecular content of the dense condensation. Local phenomena as propagation of shock waves and outflows modify molecular composition in two parts of the dense gas condensation, related to S1 and S2 YSOs.
\end{itemize}

\section*{ACKNOWLEDGEMENTS}

We are thankful to A.~O.~H.~ Olofsson for performing the observations with the APEX telescope. We also thank to A.~V.~Lapinov, A.~B.~Ostrovskii and A.~M.~Sobolev for fruitful discussions and the anonymous referee for his/her valuable comments.

This publication is based on data acquired with the Atacama Pathfinder Experiment (APEX) under programme ID 0108.F-9313(A). APEX is a collaboration between the Max-Planck-Institut fur Radioastronomie, the European Southern Observatory, and the Onsala Space Observatory. Swedish observations on APEX are supported through Swedish Research Council grant No 2017-00648.

\section*{Data Availability}
The data underlying this article are available in the ESO Archive\footnote{http://archive.eso.org/wdb/wdb/eso/apex/form} (program ID 0108.F-9313(A))

\section*{FUNDING}
K. V. Plakitina and M.~S. Kirsanova were supported by the Russian Science Foundation, grant 24-22-00097.

\section*{CONFLICT OF INTERESTS}
The authors declare no conflict of interest.








\bibliographystyle{aspb1}
\bibliography{Article}

\end{document}